\documentclass[fleqn,usenatbib]{mnras}

\usepackage{newtxtext,newtxmath}

\usepackage[T1]{fontenc}

\DeclareRobustCommand{\VAN}[3]{#2}
\let\VANthebibliography\thebibliography
\def\thebibliography{\DeclareRobustCommand{\VAN}[3]{##3}\VANthebibliography}

\usepackage{graphicx}	       
\usepackage{amsmath}	       
\usepackage{float}             
\usepackage{threeparttable}    
\usepackage{scalerel}
\usepackage{tikz}
\usetikzlibrary{svg.path}

\definecolor{orcidlogocol}{HTML}{A6CE39}
\tikzset{
  orcidlogo/.pic={
    \fill[orcidlogocol] svg{M256,128c0,70.7-57.3,128-128,128C57.3,256,0,198.7,0,128C0,57.3,57.3,0,128,0C198.7,0,256,57.3,256,128z};
    \fill[white] svg{M86.3,186.2H70.9V79.1h15.4v48.4V186.2z}
                 svg{M108.9,79.1h41.6c39.6,0,57,28.3,57,53.6c0,27.5-21.5,53.6-56.8,53.6h-41.8V79.1z M124.3,172.4h24.5c34.9,0,42.9-26.5,42.9-39.7c0-21.5-13.7-39.7-43.7-39.7h-23.7V172.4z}
                 svg{M88.7,56.8c0,5.5-4.5,10.1-10.1,10.1c-5.6,0-10.1-4.6-10.1-10.1c0-5.6,4.5-10.1,10.1-10.1C84.2,46.7,88.7,51.3,88.7,56.8z};
  }
}
\newcommand\orcidicon[1]{\href{https://orcid.org/#1}{\mbox{\scalerel*{
\begin{tikzpicture}[yscale=-1,transform shape]
\pic{orcidlogo};
\end{tikzpicture}
}{|}}}}

\title[Metallicity-dependent rates of Type Ia supernovae]{Exploring metallicity-dependent rates of Type Ia supernovae and their impact on galaxy formation}

\author[Pratik. J. Gandhi et al.]{
Pratik J. Gandhi\orcidicon{0000-0003-0965-605X}$^{1}$\thanks{E-mail: pjgandhi@ucdavis.edu; gandhipratik1995@gmail.com}, 
Andrew Wetzel\orcidicon{0000-0003-0603-8942}$^{1}$, 
Philip F. Hopkins\orcidicon{0000-0003-3729-1684}$^{2}$, 
Benjamin J. Shappee\orcidicon{0000-0003-4631-1149}$^{3}$, 
\newauthor
Coral Wheeler$^{4}$,
and Claude-André Faucher-Giguère\orcidicon{ 0000-0002-4900-6628}$^{5}$
\\
$^{1}$Department of Physics and Astronomy, University of California, Davis, CA 95616, USA\\
$^{2}$TAPIR, Mailcode 350-17, California Institute of Technology, Pasadena, CA 91125, USA\\
$^{3}$Institute for Astronomy, University of Hawaii, Honolulu, HI 96822, USA\\
$^{4}$Carnegie Observatories, Pasadena, CA 91101, USA\\
$^{5}$Department of Physics and Astronomy \& (CIERA), Northwestern University, Evanston, IL 60208, USA
}

\date{Accepted XXX. Received YYY; in original form ZZZ}

\pubyear{2021}

\begin{document}
\label{firstpage}
\pagerange{\pageref{firstpage}--\pageref{lastpage}}
\maketitle

\begin{abstract}
Type Ia supernovae play a critical role in stellar feedback and elemental enrichment in galaxies. Recent transient surveys like the All-Sky Automated Survey for Supernova (ASAS-SN) and the Dark Energy Survey (DES) find that the specific Ia rate at $z\sim0$ may be $\lesssim20-50\times$ higher in lower-mass galaxies than at Milky Way-mass. Independently, Milky Way observations show that the close-binary fraction of solar-type stars is higher at lower metallicity. Motivated by these observations, we use the FIRE-2 cosmological zoom-in simulations to explore the impact of varying Ia rate models, including metallicity dependence, on galaxies across a range of stellar masses: $10^7\,\rm{M}_{\odot}-10^{11}\,\rm{M}_{\odot}$. First, we benchmark our simulated star-formation histories (SFHs) against observations. We show that assumed SFHs and stellar mass functions play a major role in determining the degree of tension between observations and metallicity-independent Ia rate models, and potentially cause ASAS-SN and DES observations to be much more consistent with each other than might niavely appear. Models in which the Ia rate increases with decreasing metallicity (as $\propto Z^{-0.5}$ to $\propto Z^{-1}$) provide significantly better agreement with observations. Encouragingly, these increases in Ia rate ($\gtrsim10\times$ in low-mass galaxies) do not significantly impact galaxy stellar masses and morphologies: effective radii, axis ratios, and $v/\sigma$ remain largely unaffected except for our most extreme rate models. We explore implications for both [Fe/H] and [$\alpha/\rm{Fe}$] enrichment: metallicity-dependent Ia rate models can improve agreement with the observed stellar mass-metallicity relations in low-mass galaxies. Our results demonstrate that a wide range of metallicity-dependent Ia models are viable for galaxy formation and motivate future work in this area.
\end{abstract}




\section{Introduction}
\label{sec:1-intro}

Type Ia supernovae arise from the thermonuclear explosions of carbon-oxygen white dwarf (WD) stars \citep{hoyle-60} and have importance across astrophysics and cosmology. In addition to being significant sources of iron and other elements, they provide mechanical feedback in the inter-stellar medium (ISM) of galaxies - see for example \citet{iwamoto-99}, \citet{brachwitz-00}, \citet{lach-20} for discussions of Ia nucleosynthesis and \citet{matteucci-86}, \citet{kobayashi-06, kobayashi-15, kobayashi-20} for their impact on galactic and cosmic elemental enrichment. Ia are also fundamental to our study of cosmic expansion, since their intrinsic luminosities and therefore distances can be constrained with high precision \citep[][amongst others]{phillips-93, hamuy-95}, and their homogeneity allows for their use as `standard candles' -- an important rung of the extragalactic distance ladder. This makes them excellent probes of the Universe on the largest scales and of the evolution of cosmic expansion \citep[see for example][]{scolnic-13, abbott-19, riess-20}.

Despite the critical role that Ia supernovae play in astrophysics and cosmology, our understanding of their progenitors and explosion mechanisms remains limited \citep[see][for reviews on the subject]{maoz-12, wang-12, maoz-14}. There are two main competing ideas regarding the physical progenitor systems that lead to the explosion, both involving a WD in a close-binary system with a companion star. In the single-degenerate (SD) scenario, the binary companion is a main sequence (MS) star \citep[as in][]{whelan-73, nomoto-82}, while in the double-degenerate (DD) model it is another carbon-oxygen WD \citep[eg.,][]{tutukov-76, tutukov-79, iben-84, webbink-84, thompson-11, dong-15}. Both ideas have their pros and cons: although observational evidence disfavours the SD model in many cases \citep[][]{nugent-11, chomiuk-12, shappee-13, shappee-18b, tucker-20}, there are theoretical difficulties with the exact production mechanism of Ia from the DD scenario \citep[][for example]{shen-12}.

Characterizing the delay time distribution (DTD) of Ia supernovae is essential to test the validity of various progenitor scenarios. The DTD describes the supernova rate as a function of the time since star formation, and thus carries characteristic signatures of the progenitor mechanism \cite[see][for a review of observations]{wang-12, maoz-17}. The SD scenario produces a broad range of functional forms, most of which fail to account for the longer delay times seen in observations \citep[for example][]{graur-14b}. Conversely, most DD models predict a power-law DTD form of roughly $\tau^{-1}$, where $\tau$ is the delay time after the formation of a stellar population. There is also evidence that some Ia occur promptly after star formation ($\lesssim 100$ Myr), along with a delayed component that occurs much later ($\geq$ 1 Gyr) \citep{mannucci-05, scannapieco-05, sullivan-06, brandt-10, maoz-11, maoz-12}.

Another key observation in the study of Type Ia supernovae is the properties of their host galaxies. According to observations, lower-mass galaxies produce more Ia per unit stellar mass than higher-mass galaxies \citep[for example][]{mannucci-05}. Many observational surveys have studied the relationship between Ia and their host galaxies, such as the Lick Observatory Supernova Survey \citep[LOSS;][]{li-00}, the Nearby Supernova Factory \citep[SNFactory;][]{aldering-02, childress-13a}, the Texas Supernova Search \citep[TSS;][]{quimby-06}, the SuperNova Legacy Survey \citep[SNLS;][]{astier-06, guy-10}, Sloan Digital Sky Survey-II Supernova Survey \citep{frieman-08}, and the Palomar Transient Facility \citep[PTF;][]{law-09}. These efforts have identified trends between the Ia rate and host galaxy characteristics \citep[see][]{neill-06, sullivan-06, li-11a, quimby-12, smith-12, gao-13, graur-13, graur-15, graur-17, heringer-17, heringer-19}.

Recently, \citet{brown-19} showed that in the All-Sky Automated Survey for Supernova \citep[ASAS-SN;][]{shappee-13, kochanek-17}, the specific rate of Ia supernovae (total rate divided by galaxy stellar mass) is significantly higher in lower-mass galaxies, being $\sim 20 - 50$ times higher at $M_{\rm star} \sim 10^7 \, \rm{M}_{\odot}$ than at $M_{\rm star} \sim 10^{10} \, \rm{M}_{\odot}$ in their strongest observed trends. One might expect this trend \textit{qualitatively}, given that more massive galaxies have lower specific star formation rates (sSFRs) on average at $z\sim0$, and thus have lower specific Ia rates -- especially due to the strong prompt component of Ia rate models after a brief time delay following a burst of star formation. However, \citet{brown-19} find no significant difference in this mass dependence even when splitting their sample of supernova host galaxies into star-forming and quiescent. Furthermore, their inferred increase of $\sim 20 - 50\times$ appears \textit{quantitatively} too strong to be explained simply by the dependence of average sSFR on galaxy stellar mass. This seemingly indicates that the observed trend is not simply a consequence of galaxies being star-forming or quiescent at different masses, but rather has something to do with the intrinsic nature of the Ia rates themselves.

Other supernova survey results also show similar trends relative to host galaxy mass, such as \citet{graur-13} and more recently, the Dark Energy Survey \citep[DES;][]{wiseman-21}, among others. While the results from these different studies all qualitatively agree on the observed increase in Ia specific rate at lower galaxy stellar mass, they contain certain differences in the properties of their supernova and host galaxy samples. In particular, the \citet{graur-13} results span a range from $10^{10} \, \rm{M}_{\odot}$ to $10^{11.5} \, \rm{M}_{\odot}$ in galaxy stellar mass, the DES~\citep{wiseman-21} sample from $10^{8} \, \rm{M}_{\odot}$ to $10^{11.5} \, \rm{M}_{\odot}$, and the ASAS-SN~\citep{brown-19} sample covering the largest range from $10^{7} \, \rm{M}_{\odot}$ to $10^{11.5} \, \rm{M}_{\odot}$. Additionally, the volume-limited sample from ASAS-SN is the most local ($z < 0.02$), while the other two samples cover slightly higher redshift ranges while still being close to present day: $z \sim 0.1$ for \citet{graur-13} and $0.2 < z < 0.6$ for DES. Sample nuances notwithstanding, the fact remains that multiple supernova surveys have shown that the specific Ia rate increases with decreasing host galaxy mass.

In this paper we explore potential variations to models for Ia rates and DTDs that might account for this mass dependence. One motivation comes from observations of stars in the Milky Way: the close-binary fraction of solar-type stars in our Galaxy is higher in stellar populations with lower metallicity \citep[see][for example]{moe-19}. This has potential implications for Ia rates, because all widely accepted models for Ia require the presence of stars in a close-binary system. To the extent that stellar populations in lower-mass galaxies are more metal-poor on average, \citep[for example][]{lequeux-79, tremonti-04, gallazzi-05, lee-06, kirby-13, leethochawalit-18}, lower-mass galaxies should have higher close-binary fractions, potentially boosting their Ia rates per unit stellar mass relative to higher-mass galaxies. For a related viewpoint regarding the dependence of wide binary fractions on iron abundance, see \citet{hwang-21}.

Further motivation for metallicity dependence to Ia rates also comes from other observations of Milky Way stellar populations: blue straggler stars (BSS) are more common at lower metallicities \citep[as discussed in][]{wyse-20}. While this study argues that mass transfer in close-binary pairs is the likely cause for BSS formation, this result has implications for Ia rates too, by lending further credence to the prevalence of higher close-binary fractions in metal-poor populations. This motivates our exploration of metallicity-dependence Ia rate models in an effort to investigate the mass dependence of specific Ia rates, and their implications for the astrophysics of stellar feedback and elemental enrichment in galaxy formation.

A number of previous studies have examined different models for the rates of core-collapse supernova in galaxy simulations \citep{springel-03, agertz-13, crain-15, kim-16, rosdahl-17, peeples-19, keller-20}, and certain previous observational studies \citep[such as][]{cooper-09} have discussed metallicity dependence of observed Ia rates. \textit{However, to our knowledge our paper is the first to explore  metallicity-dependent models for Ia rates using cosmological zoom-in simulations, through which we also study their impact on galaxy formation.} 


\section{Methods}
\label{sec:2-methods}

\begin{table}
\centering
\caption{
    \textbf{The properties of the FIRE-2 simulations that we analyse at $z = 0$.}
    These are the original FIRE-2 simulations that used the \citet{mannucci-06} DTD, and form the fiducial suite that we use for our analysis. For details of the re-simulations with varying rate models, refer to Table \ref{tab:modified-sims} and Section~\ref{sec:2.4-ia-rates-met}. We measure all properties (including total stellar mass and Ia rates) using all star particles within a $15$ kpc spherical volume around each host galaxy. We list the initial star and gas particle masses under `Baryonic Resolution'. $\tau_{90}$ denotes the lookback time from present day at which a galaxy assembled $90$ per cent of its current stellar mass, while sSFR denotes the specific star formation rate at $z\sim0$ (averaged over the last $500$ Myr in lookback time).
    }
    \addtolength{\tabcolsep}{-2pt}
    \begin{tabular}{|c|c|c|c|c|c|}
	    \hline
	    \hline
	    Name$^{\dagger}$ & Stellar & Baryonic & $\tau_{90}$ & sSFR & Ref.\\
	    & Mass & Resolution & (lookback) & $[10^{-11}]$ &\\
	    & [$\rm{M}_{\odot}$] & [$\rm{M}_{\odot}$] & [Gyr] & $[\rm{yr}^{-1}]$ &\\
	    \hline
	    m12m & $1.2\times10^{11}$ & $7100$ & $1.49$ & $11.2$ & A\\
	    Romulus & $9.3\times10^{10}$ & $4000$ & $1.38$ & $13.0$ & B\\
	    m12b & $9.1\times10^{10}$ & $7100$ & $1.27$ & $12.0$ & C\\
	    m12f & $8.2\times10^{10}$ & $7100$ & $1.19$ & $11.4$ & D\\
	    Thelma & $7.5\times10^{10}$ & $4000$ & $1.08$ & $14.8$ & C\\
	    Romeo & $6.9\times10^{10}$ & $3500$ & $2.13$ & $8.10$ & C\\
	    m12i & $6.8\times10^{10}$ & $7100$ & $1.45$ & $11.1$ & E\\
	    m12c & $6.3\times10^{10}$ & $7100$ & $0.98$ & $15.6$ & C\\
	    m12w & $6.1\times10^{10}$ & $7100$ & $0.92$ & $22.7$ & F\\
	    m11g & $5.2\times10^{10}$ & $12000$ & $0.97$ & $18.6$ & G\\
	    Remus & $4.8\times10^{10}$ & $4000$ & $1.89$ & $8.34$ & B\\
	    Juliet & $4.1\times10^{10}$ & $3500$ & $1.57$ & $10.3$ & C\\
	    Louise & $2.7\times10^{10}$ & $4000$ & $1.52$ & $10.5$ & C\\
	    m11f & $2.7\times10^{10}$ & $12000$ & $0.76$ & $19.1$ & G\\
	    m12z & $2.2\times10^{10}$ & $4200$ & $0.55$ & $25.3$ & C\\
	    m12r & $1.8\times10^{10}$ & $7100$ & $0.83$ & $15.7$ & F\\
	    m11d & $4.0\times10^{9}$ & $7100$ & $0.78$ & $18.0$ & H\\
	    m11e & $1.5\times10^{9}$ & $7100$ & $0.90$ & $16.7$ & H\\
	    m11v & $1.1\times10^{9}$ & $7100$ & $0.62$ & $21.6$ & A\\
	    m11i & $1.0\times10^{9}$ & $7100$ & $1.14$ & $9.07$ & H\\
	    m11c & $8.9\times10^{8}$ & $2100$ & $0.91$ & $16.7$ & A\\
	    m11q & $4.0\times10^{8}$ & $880$ & $1.65$ & $7.69$ & A\\
	    m11h & $1.4\times10^{8}$ & $880$ & $3.02$ & $1.01$ & H\\
	    m11a & $1.3\times10^{8}$ & $2100$ & $1.30$ & $15.2$ & G\\
	    m11b & $4.9\times10^{7}$ & $2100$ & $2.17$ & $1.68$ & G\\
	    m10z & $4.0\times10^{7}$ & $250$ & $0.93$ & $7.03$ & A\\
	    m10y & $2.3\times10^{7}$ & $250$ & $2.06$ & $2.50$ & A\\
	    m09 & $1.2\times10^{4}$ & $250$ [\& $30$] & $11.7$ & $\approx0$ & I\\ 
	    \hline
	    \hline
\end{tabular}
\addtolength{\tabcolsep}{2pt}
\label{tab:FIRE-2-sims}
\begin{tablenotes}
\item $^{\dagger}$Simulation first introduced at this resolution in: A: \citet{hopkins-18-FIRE}, B: \citet{garrison-kimmel-19b}, C: \citet{garrison-kimmel-19a}, D: \citet{garrison-kimmel-17}, E: \citet{wetzel-16}, F: \citet{samuel-20}, G: \citet{chan-18}, H: \citet{el-badry-18}, and I: \citet{wheeler-19}.
\end{tablenotes}
\end{table}

\subsection{FIRE-2 simulations}
\label{sec:2.1-sims-intro}

We use a fiducial suite of 28 cosmological zoom-in simulations of galaxies with stellar masses from $10^7 \, \rm{M}_{\odot}$ - $10^{11} \, \rm{M}_{\odot}$, from the Feedback In Realistic Environments (FIRE) project\footnote{\url{http://fire.northwestern.edu}} \citep{hopkins-18-FIRE}. FIRE simulations are run using \textsc{Gizmo}, a Lagrangian Meshless Finite Mass (MFM) hydrodynamics code \citep{hopkins-15}. Each simulation includes an implementation of fluid dynamics, star formation, and stellar feedback based on the FIRE-2 numerical prescription. FIRE-2 models the multi-phase inter-stellar medium (ISM) in galaxies and incorporates physically motivated, metallicity-dependent radiative heating and cooling processes for gas. These include free-free, photoionisation and recombination, Compton, photo-electric and dust collisional, cosmic ray, molecular, metal-line, and fine structure processes. They account for $11$ elements (H, He, C, N, O, Ne, Mg, Si, S, Ca, Fe) across a temperature range of $10 - 10^{10} \rm{K}$. They also include the sub-grid diffusion and mixing of these elements in gas phase via turbulence \citep[see][for further details]{escala-18, hopkins-18-FIRE}. The fiducial FIRE-2 simulations also include a spatially uniform, redshift-dependent UV background from \citet{faucher-giguere-09}. In calculating metallicities throughout this paper, we scale elemental abundances to the (proto-)Solar values from \citet{asplund-09}.

Star particles form out of gas that is self-gravitating, Jeans-unstable, cold ($T < 10^{4} \, \rm{K}$), dense ($n > 10^3 \, \rm{cm}^{-3}$), and molecular \citep[following][]{krumholz-11}. Each star particle represents a single stellar population, assuming a \citet{kroupa-01} stellar initial mass function. During formation, star particles also inherit the mass and elemental abundances of their respective progenitor gas particles. In FIRE-2, star particles evolve along standard stellar population models from e.g. STARBURST99 v7.0 \citep{leitherer-99}. We also include the following time-resolved stellar feedback processes: core-collapse and Type Ia supernovae, continuous mass loss, radiation pressure, photoionisation, and photo-electric heating. FIRE-2 uses rates for core-collapse supernovae from STARBURST99 \citep{leitherer-99}, and their nucleosynthetic yields from \citet{nomoto-06}. Stellar wind yields, sourced primarily from O, B, and AGB stars, are from a combination of models from \citet{van-den-hoek-97}, \citet{marigo-01}, and \citet{izzard-04}, synthesised in \citet{wiersma-09}.
For details on the implementation of Ia supernovae, see Section~\ref{sec:2.3-ia-rates}. For a discussion of the implementation of supernova feedback energetics in FIRE-2, see \citet{hopkins-18-SN}.

We generate cosmological zoom-in initial conditions for each simulation at $z = 99$ using the \textsc{MUSIC} code \citep{hahn-11}. These initial conditions are embedded within periodic cosmological boxes of side length ranging from $70$ to $172$ Mpc. We save 600 snapshots per simulation from $z=99$ to $z=0$, with an average spacing of $\lesssim 25$ Myr. For all simulations we assume flat $\Lambda$CDM cosmology, using parameters broadly consistent with those from \citet{planck-18}: $h = 0.68-0.71$, $\Omega_\Lambda = 0.69 - 0.734$, $\Omega_{\rm m} = 0.266 - 0.31$, $\Omega_{\rm b} = 0.0455 - 0.048$, $\sigma_8 = 0.801 - 0.82$, and $n_{\rm s} = 0.961 - 0.97$.

In each simulation, we consider only the host galaxy and not the satellites. Our sample of simulated galaxies contains 15 hosts with dark matter (DM) halo masses of $M_{200\rm{m}} \approx 10^{12} \, \rm{M}_{\odot}$ (labelled as `m12', along with `Romeo', `Juliet', `Romulus', `Remus', `Thelma', and `Louise'), 10 with halo masses of $M_{200\rm{m}} \sim 10^{11} \, \rm{M}_{\odot}$ (labelled as `m11'), 2 with halo masses of $M_{200\rm{m}} \sim 10^{10} \, \rm{M}_{\odot}$ (labelled as `m10'), and 1 with a halo mass of $M_{200\rm{m}} \sim 10^{9} \, \rm{M}_{\odot}$ (labelled as `m09'). Here $M_{200\rm{m}}$ refers to the total mass within the radius containing 200 times the mean matter density of the Universe.

Table~\ref{tab:FIRE-2-sims} lists all of the (fiducial) FIRE-2 simulations that we use along with the stellar masses of each galaxy, the initial baryonic particle masses in the simulation, the lookback time to when each galaxy assembled $90$ per cent of its stellar mass at $z = 0$, and its sSFR at $z \approx 0$.

In this paper, we examine not only metallicity-dependent models for Ia supernova rates using the FIRE simulations, but also their dependence on and impact on various galaxy properties -- star formation rate (SFR), star formation history (SFH), morphology, and elemental abundance. Previous studies have benchmarked a number of these properties for the fiducial FIRE-2 simulations, and we list them here for reference. \citet{hopkins-18-SN} describe the implementation and testing of supernova modelling in FIRE-2, but without much specific detail on different models for Ia rates. For discussion of stellar mass assembly timescales and SFHs in FIRE-2 galaxies, see \citet{garrison-kimmel-19a}, \citet{garrison-kimmel-19b}, \citet{graus-19}, \citet{iyer-20}, and \citet{santistevan-20}. \citet{sparre-17} and \citet{flores-velasquez-21} examined SFRs of FIRE-2 galaxies using different observational tracers, while \citet{emami-21} tested the impact of bursty SFHs on the sizes of FIRE-2 low-mass galaxies. For stellar elemental abundances in FIRE, \citet{ma-16}, \citet{escala-18}, and \citet{hopkins-20} benchmarked both iron and alpha-capture element distributions in high-mass and low-mass galaxies, while \citet{wheeler-19} provided the same for ultra-faint galaxies.

\subsection{Benchmarking star-formation histories in FIRE-2}
\label{sec:2.2-sfh-benchmarking}

Before studying supernova rates, we first examine the SFHs for FIRE-2 galaxies. Irrespective of choice of Ia DTD (either the \citet{mannucci-06} or the \citet{maoz-17} models), at any given time, Ia rates should be most sensitive to recent star formation. In Appendix~\ref{app:A} and Figure~\ref{fig:rate-lbt}, we show that for the entire range in galaxy masses we consider, star formation within the last $\sim 1 \rm{Gyr}$ in lookback time accounts for the majority of Ia events at $z = 0$, with lower-mass galaxies being somewhat more sensitive to slightly earlier star formation. Thus, we benchmark recent SFHs in our simulations against a compilation of observations and semi-empirical models of galaxies from $10^7 \, \rm{M}_{\odot}$ to $10^{11} \, \rm{M}_{\odot}$, to provide confidence in using FIRE-2 to compare to observed Ia rates and to test alternative rate models.

\begin{figure*}
\centering
\begin{tabular}{c}
\includegraphics[width =  1.0\linewidth]{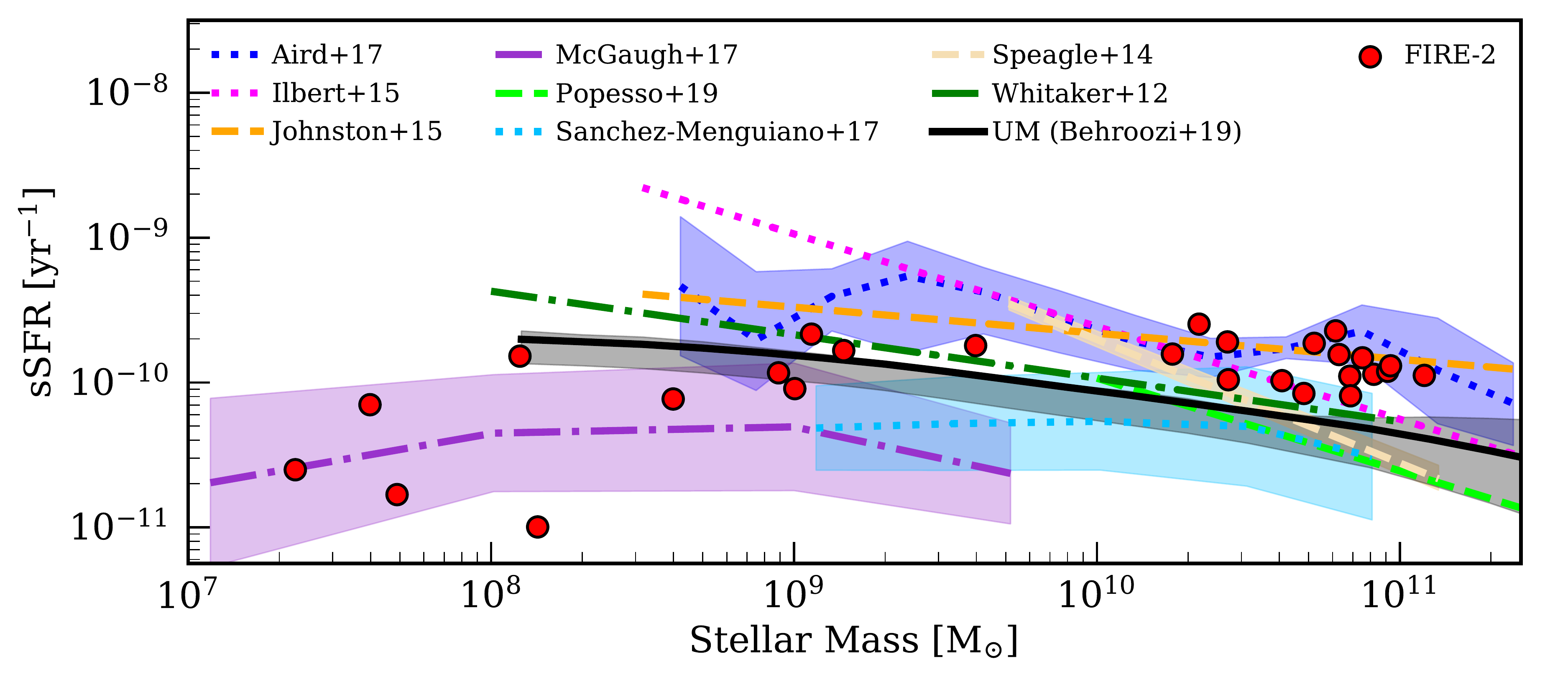}
\end{tabular}
\vspace{-2 mm}
\caption{
\textbf{Specific star formation rate (sSFR) at $z = 0$.} We show sSFR averaged over the last $500$ Myr for FIRE-2 galaxies (red points), compared to observations \citep{whitaker-12, speagle-14, ilbert-15, johnston-15, aird-17, mcgaugh-17, popesso-19, sanchez-menguiano-19}, which we show as best-fit lines with (where provided) $1-\sigma$ scatter. We also compare to sSFR values from the semi-empirical model \textsc{UniverseMachine} \citep[UM;][]{behroozi-19}. Across our entire mass range, FIRE-2 sSFRs fall within the range of observed values, although they do not necessarily agree with a single observational fit extrapolated across the entire mass range. There is general agreement even at for our lower-mass simulations, as also discussed in Appendix \ref{app:B} and Figure \ref{fig:tau-90}. \textit{Because the specific rates of Ia supernovae at $z \sim 0$ are primarily sensitive to very recent star formation (within $\lesssim 1$ Gyr) and thus to sSFRs at $z \sim 0$, this broad agreement, within significant observational scatter, provides an important benchmark of our simulated SFHs and resultant Ia rates.}
}
\label{fig:SFMS-z0}
\end{figure*}

Figure~\ref{fig:SFMS-z0} shows the sSFR at $z = 0$ versus stellar mass for each FIRE-2 galaxy. We also show sSFRs from a compilation of observations as well as from the semi-empirical \textsc{UniverseMachine} model \citep[hereafter UM;][]{behroozi-19}. In computing present-day sSFRs, we take the mean of the sSFR, $\langle \dot{M}_* \rangle$, over the last $500 \rm{Myr}$ in lookback time, to account for stochasticity in the simulations and to better match the redshift range of the observations.  We show the mean or median trend line along with $68$ per cent confidence intervals for the observed samples wherever applicable, while for the remaining observations we provide a best-fit relation to compare against.

The intrinsic scatter between different observational star-forming main sequence (SFMS) relations is large, likely from the variety of techniques used to infer SFRs in their respective samples. \citet{whitaker-12} rely on \textit{Spitzer}-MIPS fluxes from the S-COSMOS and FIDEL surveys. \citet{speagle-14} present a compilation of SFMS measurements from a variety of observational papers using different techniques for measuring SFRs. \citet{ilbert-15} rely on mid- and far-infrared observations of a catalogue of galaxies from the COSMOS and GOODS surveys. \citet{johnston-15} employ SED fitting of broad-band photometry and mid-infrared data from the VIDEO survey. \citet{aird-17} use deep \textit{Chandra} X-ray luminosity observations and measurements of high- and low-mass X-ray binary stars. \citet{mcgaugh-17} estimate SFRs for lower-mass (and low-surface-brightness) galaxies using a compilation of H$\alpha$ flux measurements. \citet{popesso-19} leverage infrared observations from WISE and \textit{Herschel} as well as H$\alpha$ fluxes. Finally, \citet{sanchez-menguiano-19} derive SFRs from dust-corrected H$\alpha$ luminosities of galaxies in the SDSS-IV MaNGA survey. We do not claim to prefer any one sample over the others, because the differences in the various observational techniques are beyond the scope of this paper - for a more detailed study of different SFR tracers in the FIRE simulations, see \citet{flores-velasquez-21} instead. Here, however, we present the entire spread of observed samples to compare our simulations against.

Nominally, the FIRE-2 trendline for all $M_*$ does not necessarily agree with \textit{any single observed sample extrapolated across the entire mass range} -- in particular, some such as \citet{ilbert-15} show a significantly steeper trend in sSFR. However, all of our $M_* \gtrsim 10^9 \, \rm{M}_{\odot}$ simulations broadly agree with the full range of observed sSFRs after accounting for $1\sigma$ uncertainties and systematic scatter between datasets. Although extrapolations of the higher-mass observed SFMS to $M_* < 10^9 \, \rm{M}_{\odot}$ would suggest tension with our lower-mass simulations, those simulations agree within $1 \sigma$ with the \citet{mcgaugh-17} values, measured directly for low-mass galaxies. Additionally, in Appendix~\ref{app:B} and Figure~\ref{fig:tau-90}, we compare the $90$ per cent stellar mass assembly timescales for FIRE-2 galaxies with observed values for Local Group galaxies as well as observed and semi-empirical values at high masses. That comparison shows good systematic agreement between our low-mass simulations and Local Group observations. Since a slightly longer star formation timescale is relevant for $z\sim0$ Ia rates for low-mass galaxies, this comparison provides confidence in our low-mass SFHs despite the seeming tension with extrapolations of the high-mass SFMs. Altogether, examining our simulated SFHs validates their accuracy, especially in the context of their Ia rates being `realistic' and not biased by the idiosyncrasies of their recent SFHs.

\subsection{Type Ia supernovae: delay time distributions and nucleosynthetic yields}
\label{sec:2.3-ia-rates}

\begin{figure*}
\centering
\begin{tabular}{c c c}
\includegraphics[width=0.32\linewidth]{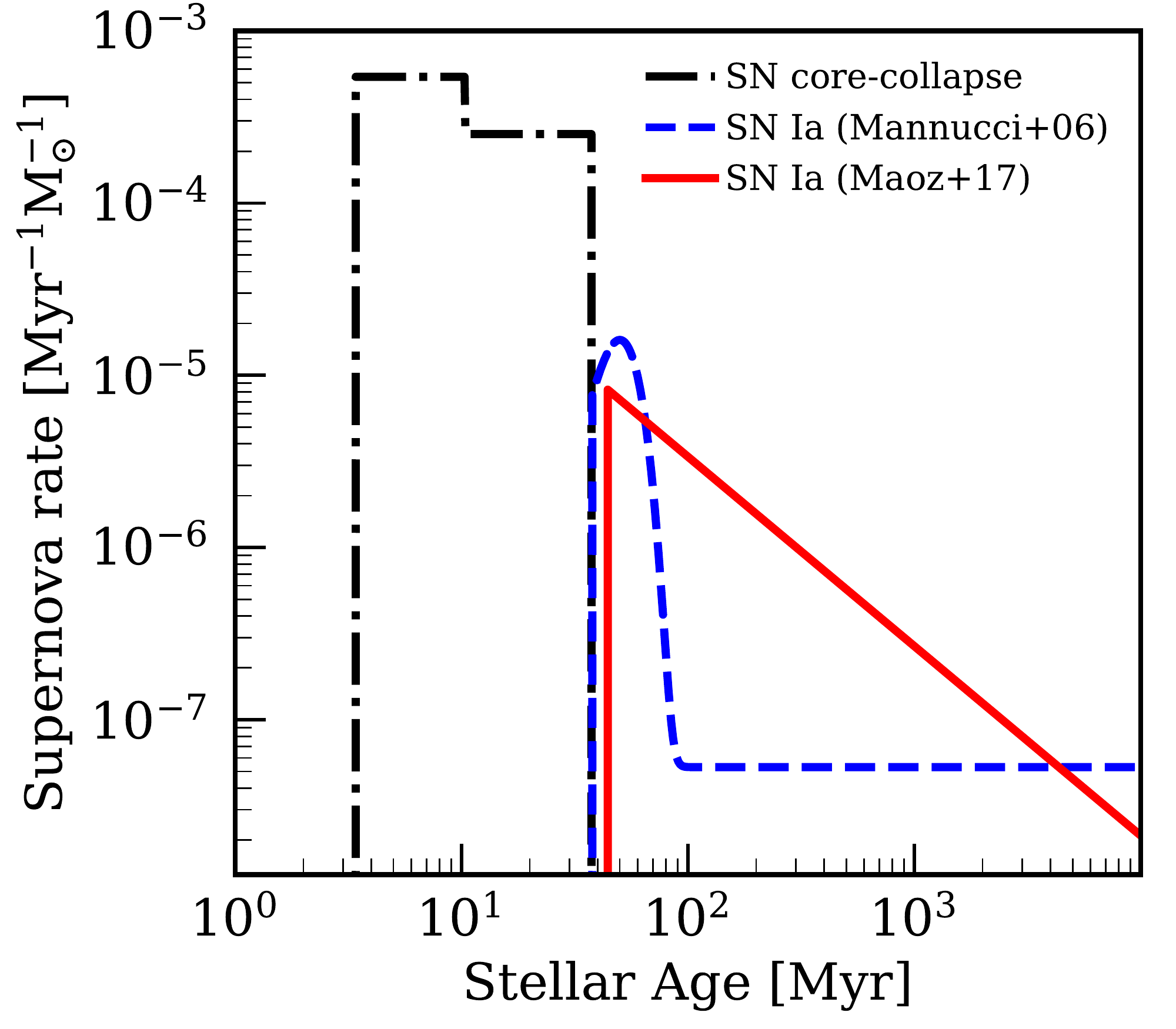} &
\includegraphics[width=0.32\linewidth]{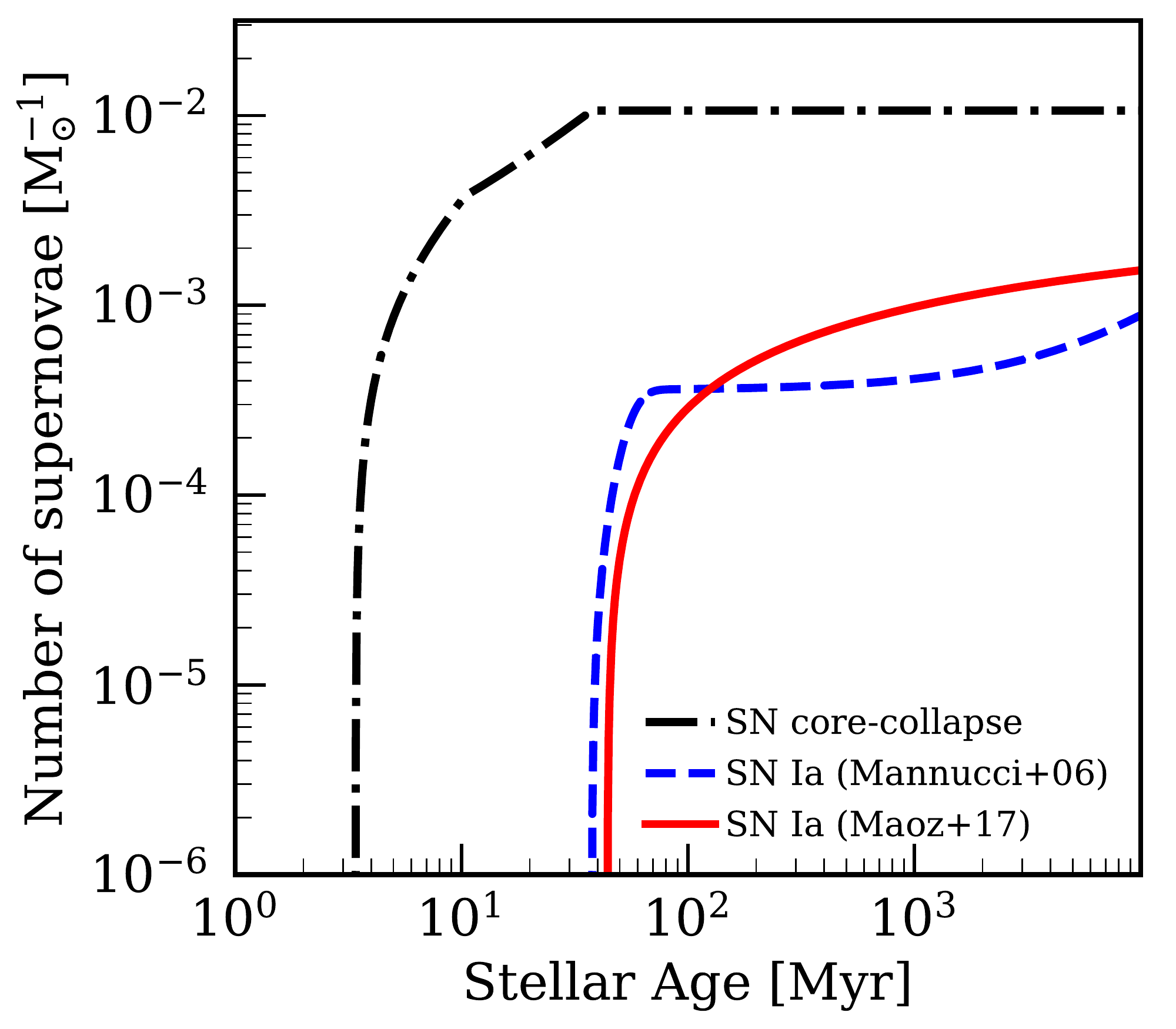} &
\includegraphics[width=0.32\linewidth]{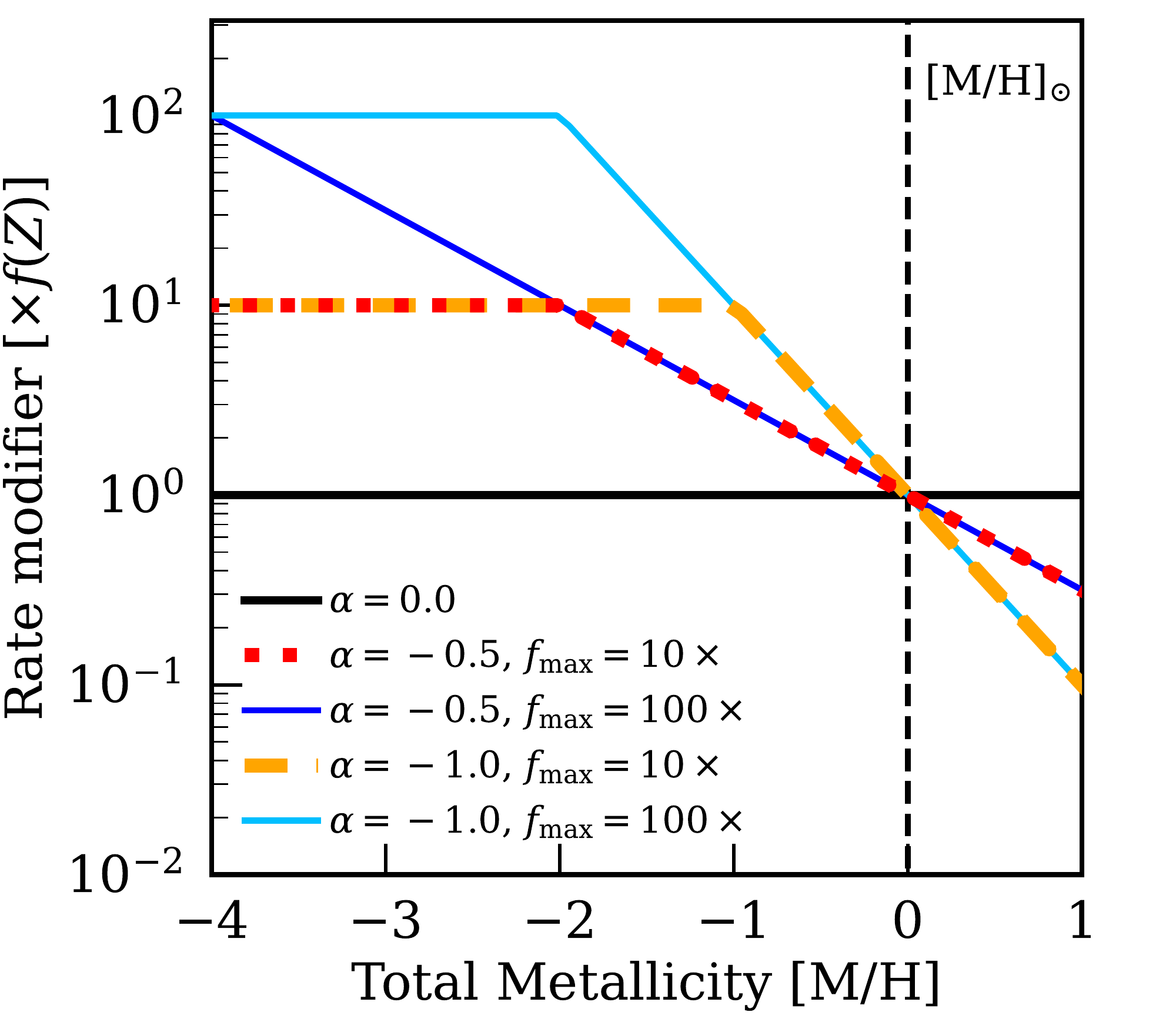}
\end{tabular}
\caption{
\textbf{Comparison of models for the delay time distribution (DTD) of Type Ia supernovae. Left \& Centre:} specific rate (rate divided by stellar mass; left) and cumulative specific number (centre) of Ia versus stellar age.
\citet{mannucci-06} (blue) is the fiducial DTD used in FIRE-2 simulations, while the \citet{maoz-17} DTD (red) is what we use for all analysis and re-simulation in this work. The ratio of the cumulative number of Ia supernovae in \citet{maoz-17} relative to \citet{mannucci-06} is: $2.4$ at $1$ Gyr, $1.81$ at $10$ Gyr, and $1.47$ at $13.7$ Gyr. For reference, we also show the DTD of choice in FIRE-2 for core-collapse supernovae in black \citep[as per][]{leitherer-99}. The cumulative number of core-collapse is $\sim 10 \times$ that of Ia, so Ia do not dominate feedback energetics unless boosted by $\gtrsim 10 \times$. \textbf{Right}: Models for our metallicity-dependent Ia rate modifier to the \citet{maoz-17} DTD, where the rate normalisation is multiplied by a power law based on total metallicity of a star particle \citep[normalised to Solar as per][]{asplund-09} raised to a negative exponent, along with a choice of an artificial rate boost cap, $f_{\rm max}$. These models simply multiply the normalisation of the Ia rates; they do not change the shape of the DTD. See Section~\ref{sec:2.4-ia-rates-met} and Equation~\ref{eq:met-modifier} for more on our metallicity-dependent modifiers.
}
\label{fig:SN-DTD}
\end{figure*}

FIRE-2 assumes nucleosynthetic yields for Type Ia supernovae from \citet{iwamoto-99}. The original Ia DTD implementationin FIRE-2 is that of \citet{mannucci-06}, which uses a prompt Gaussian component followed by a constant rate at later times. However, the DTD from \citet{maoz-17}, with a roughly $\tau^{-1}$ form, is more observationally and theoretically motivated. Thus, for our analysis here, we focus on the DTD from \citet{maoz-17} and metallicity-dependent modifications to its normalisation (as described in Section~\ref{sec:2.4-ia-rates-met}). We also examine re-simulations of specific galaxies with different metallicity-dependent Ia rate models, as discussed in Section~\ref{sec:2.4-ia-rates-met} and Table~\ref{tab:modified-sims}. The left and centre panels of Figure~\ref{fig:SN-DTD} compare the functional forms of the two DTDs, with the FIRE-2 implementation of the core-collapse supernova rates also shown for comparison.

As Figure~\ref{fig:SN-DTD} (centre) shows, the \citet{maoz-17} DTD leads to a higher cumulative number of Ia for times $\geq 100$ Myr, and the ratio of the number of supernovae relative to the \citet{mannucci-06} DTD is: $2.4$ at $1$ Gyr, $1.81$ at $10$ Gyr, and $1.47$ at $13.7$ Gyr. Also, the total number of core-collapse supernovae is $\sim10\times$ higher than Ia in either case, and the mechanical kinetic energy of IMF-averaged supernova events implemented in FIRE-2 is identical for core-collapse and Ia ($\sim 10^{51}$ erg), so Ia do not dominate supernova feedback unless their rate is increased by more than $10\times$, which is important for understanding the results from our metallicity-dependent Ia rates. To explore the overall normalization of this rate increase, we also impose an (arbitrary) cap on how much the rate can be boosted -- either $10\times$ or $100\times$, as Equation~\ref{eq:met-modifier} shows.

During simulation runtime, \textsc{Gizmo} samples supernova events probabilistically from the DTD based on the age of each star particle. For the first part of our analysis, we re-compute Ia rates keeping galaxy properties fixed (as shown in Table~\ref{tab:FIRE-2-sims}), by post-processing a single $z=0$ snapshot. We consider the ages (at $z=0$) of all star particles within a $15$ kpc spherical volume around each host galaxy, and we apply them to the DTD of choice to retrieve a Ia rate for each star particle. We then add these to obtain an overall Ia rate for the galaxy, and we divided by the total stellar mass of the galaxy at $z = 0$ to compute the specific Ia rate. To account for stochasticity (burstiness) in the star-formation rates, and to roughly match the redshift distribution of the observations that we compare against, we average the Ia rate over the past $500$ Myr in lookback time from $z=0$.

\subsection{Metallicity-dependent Ia rates and re-simulations}
\label{sec:2.4-ia-rates-met}

Motivated by observations of the Milky Way showing that the close-binary fraction of stars is higher in lower-metallicity stars (see Section~\ref{sec:1-intro}), we introduce a power-law metallicity-dependent modifier to the normalization of the Ia DTD from \citet{maoz-17}. Note that a detailed theoretical examination of the physics behind such a metallicity dependence and its exact nature are beyond the scope of this paper; instead we carry out an empirical analysis of metallicity-dependent rate modifications using the FIRE-2 simulations, and their macroscopic effects on stellar feedback, elemental enrichment, and galaxy formation in general. 

Our modifications do not change the shape of the DTD nor the minimum delay time: we only modify its normalisation based on the total metal mass fraction of the star particle for which the rate is being computed, as shown below.

\begin{equation}
    \frac{dN_{\rm Ia}}{dtdM_*} \equiv \frac{dN_{\rm Ia}}{dtdM_*}\bigg|_{0} \times f(Z) \; \; \rm{where} \; \; \textit{f(Z)} \equiv \rm{min} \left[ \left(\frac{\textit{Z}}{\textit{Z}_{\odot}}\right)^{\alpha} , \; \; \textit{f}_{\rm max} \right]
    \label{eq:met-modifier}
\end{equation} \newline

Equation~\ref{eq:met-modifier} shows the functional form for our power-law modifier, where for each star particle, we multiply the normalisation of the \citet{maoz-17} DTD (denoted by $dN_{\rm Ia}/dtdM_*|_0$) with the minimum of the following two quantities: (a) a metallicity-dependent power law, normalised to the solar metal mass fraction \citep[from][]{asplund-09}, with a variable exponent (denoted by $\left(Z/Z_{\odot}\right)^{\alpha}$), and (b) an (arbitrary) cap $f_{\rm max}$ of either $10\times$ or $100\times$ imposed on the rate enhancement. The latter allows us to control the strength of the total boost to the Ia rate and ensure that it is not arbitrarily large for populations with arbitrarily low metallicities. A cap of $10 \times$ also means that the feedback from Ia is comparable to (does not dominate over) that from core-collapse supernovae (which we do not modify) at low metallicities.

Figure~\ref{fig:SN-DTD} (right) provides a visual representation of the metallicity dependence of our modifier, including the normalisation to Solar abundance. The impact of the choice of rate-boost cap also is visible for [M/H] $ \leq -2$. Again, the (simple) modifiers that we explore affect only the normalisation of the \citet{maoz-17} DTD, without altering its shape.

There is an important distinction between re-computing rates in post-processing with our metallicity-dependent models and fixed galaxy properties, and self-consistently re-simulating galaxies with modified Ia rate models. To test our metallicity-dependent rate models and compare to results from observational surveys, we first compute Ia rates for various FIRE-2 simulations by applying these models in post-processing only. While this allows us to estimate the Ia rates resulting from various modifications to the DTD, it does not change the simulation's feedback energetics, enrichment, or SFH \citep[all of which are based on the original simulations using the DTD from][]{mannucci-06}. Note that this is not an issue for the fiducial, metallicity-independent \citet{maoz-17} DTD, since the difference in the number of Ia over $\sim10$ Gyr is $\sim50$ per cent, but becomes a major concern when we introduce metallicity dependence. Therefore, to self-consistently model the simulations' formation histories, we re-simulate a subset (4) of the FIRE-2 galaxies (m12i, m11e, m11b and m09, with original stellar masses of $\sim 10^{10} \, \rm{M_{\odot}}$, $10^{8} \, \rm{M_{\odot}}$, $10^{7} \, \rm{M_{\odot}}$, and $10^{4} \, \rm{M_{\odot}}$ respectively) using the DTD from \citet{maoz-17} and various metallicity-dependent rate modifiers. We then examine their (self-consistent) Ia rates, elemental abundances, and other galaxy properties. Table~\ref{tab:modified-sims} lists these re-simulations, including the input models that we use and the resultant galaxy properties at $z=0$.

\begin{table*}
\centering
\caption
    {\textbf{Re-simulations with varying Ia rate models}. All analysis (including total stellar mass, Ia rates, and elemental abundances) is at $z=0$ using star particles within $15$ kpc from the galaxy's centre. Baryonic resolutions (initial star/gas particle masses) for each set of re-simulations are: $57000\,\rm{M}_{\odot}$ for all m12i, $7100\,\rm{M}_{\odot}$ for all m11e, $2100\,\rm{M}_{\odot}$ for all m11b, $250\,\rm{M}_{\odot}$ for m09Mann and m09Z05-100x, and $30\,\rm{M}_{\odot}$ for m09Maoz and m09Z05-10x. For each one, we also show the choices of power-law exponent, or $\alpha$, and rate boost cap, or $f_{\rm max}$ (as described in Equation~\ref{eq:met-modifier}). $\tau_{90}$ denotes the time at which a galaxy assembled $90$ per cent of its current stellar mass, while sSFR denotes the $z\sim0$ specific star formation rate (those with sSFR $\approx 0$ quenched before present day from increased Ia supernova feedback, or in the case of m09 and its re-simulations, reionisation). For m12i re-simulations, [Fe/H] and [Mg/Fe] represent $M_*$-weighted linear mean values for all selected star particles, while for m11e and m11b re-simulations, they represent $M_*$-weighted median values (as described in Section~\ref{sec:3.2.3-MZR}, Equation~\ref{eq:met-SMWLM}, and Figure~\ref{fig:MZR-massive}). For each m09 re-simulation, we show the [Fe/H] and [Mg/Fe] values that represent the $M_*$-weighted median for only those star particles that have been enriched and are no longer at the imposed initial metallicity floor (further details in Section~\ref{sec:3.2.4-FeH-UFD} and Figure~\ref{fig:MZR-UFD}).
    } 
    \begin{tabular}{ccccccccccc}
	    \hline
	    \hline
	    Name & Power-law & Rate & Stellar & $\tau_{90}$ & sSFR & [Fe/H] & [Mg/Fe] & $R_{90}$ & $v/\sigma$ & Ax. ratio\\
	    & Exponent & Boost Cap & Mass & (lookback) & $[\times 10^{-11}]$ & (stellar) & (stellar) & (stellar) & (stellar) & (stellar)\\
	    & ($\alpha$) & ($f_{\rm max}$) & [$\rm{M}_{\odot}$] & [Gyr] & [$\rm{yr}^{-1}$] &  & & [kpc] & & (min/maj) \\
		\hline
        m12i\_Mann$^{\dagger}$ & N/A & N/A & $1.4 \times10^{11}$ & $2.75$ & $5.64$ & $0.05$ & $0.27$ & $7.1$ & $1.6$ & $0.12$\\
        m12i\_Maoz & N/A & N/A & $1.4\times10^{11}$ & $2.73$ & $6.18$ & $0.21$ & $0.11$ & $6.0$ & $1.6$ & $0.15$\\
        m12i\_Z05-10x & $-0.5$ & $10\times$ & $1.4\times10^{11}$ & $2.45$ & $7.05$ & $0.13$ & $0.12$ & $7.5$ & $1.7$ & $0.12$\\
        m12i\_Z05-100x & $-0.5$ & $100\times$ & $1.3\times10^{11}$ & $2.36$ & $6.21$ & $0.12$ & $0.12$ & $7.7$ & $1.8$ & $0.11$\\
        m12i\_Z1-10x & $-1.0$ & $10\times$ & $1.3\times10^{11}$ & $1.72$ & $9.27$ & $0.09$ & $0.10$ & $9.0$ & $1.7$ & $0.13$\\
        m12i\_Z1-100x & $-1.0$ & $100\times$ & $9.1\times10^{10}$ & $1.03$ & $16.1$ & $0.08$ & $0.06$ & $10.1$ & $1.6$ & $0.12$\\
        \hline
        m11e\_Mann$^{\dagger}$ & N/A & N/A & $4.4\times10^{8}$ & $0.96$ & $14.4$ & $-1.37$ & $0.24$ & $9.4$ & $0.2$ & $0.26$\\
        m11e\_Maoz & N/A & N/A & $6.2\times10^{8}$ & $0.84$ & $14.8$ & $-0.89$ & $0.05$ & $8.8$ & $0.2$ & $0.32$\\
        m11e\_Z05-10x & $-0.5$ & $10\times$ & $3.2\times10^{8}$ & $1.27$ & $11.3$ & $-0.68$ & $-0.34$ & $14.8$ & $0.3$ & $0.44$\\
        m11e\_Z05-100x & $-0.5$ & $100\times$ & $2.8\times10^{8}$ & $1.31$ & $10.6$ & $-0.79$ & $-0.42$ & $10.1$ & $0.1$ & $0.28$\\
        m11e\_Z1-10x & $-1.0$ & $10\times$ & $9.7\times10^{7}$ & $3.00$ & $6.35$ & $-1.07$ & $-0.63$ & $17.8$ & $0.3$ & $0.53$\\
        m11e\_Z1-100x & $-1.0$ & $100\times$ & $2.3\times10^{7}$ & $10.84$ & $\approx0$ & $-1.18$ & $-0.78$ & $15.7$ & $0.2$ & $0.64$\\
        \hline
        m11b\_Mann$^{\dagger}$ & N/A & N/A & $4.9\times10^{7}$ & $4.69$ & $3.13$ & $-1.87$ & $0.23$ & $6.3$ & $0.6$ & $0.31$\\
        m11b\_Maoz & N/A & N/A & $6.2\times10^{7}$ & $4.78$ & $1.88$ & $-1.68$ & $0.07$ & $6.7$ & $0.4$ & $0.31$ \\
        m11b\_Z05-10x & $-0.5$ & $10\times$ & $3.3\times10^{7}$ & $5.68$ & $2.78$ & $-1.33$ & $-0.45$ & $5.8$ & $0.6$ & $0.35$ \\
        m11b\_Z05-100x & $-0.5$ & $100\times$ & $3.4\times10^{7}$ & $0.33$ & $25.9$ & $-1.09$ & $-0.57$ & $6.4$ & $1.2$ & $0.15$ \\
        m11b\_Z1-10x & $-1.0$ & $10\times$ & $3.2\times10^{7}$ & $3.85$ & $\approx0$ & $-1.17$ & $-0.63$ & $6.8$ & $0.6$ & $0.29$ \\
        m11b\_Z1-100x & $-1.0$ & $100\times$ & $2.5\times10^{6}$ & $2.35$ & $\approx0$ & $-1.57$ & $-0.89$ & $10.8$ & $0.4$ & $0.76$ \\
        \hline
        m09\_Mann$^{\dagger}$ & N/A & N/A & $5.8\times10^{4}$ & $11.78$ & $\approx0$ & $-3.40$ & $0.18$ & $6.0$ & $0.6$ & $0.34$ \\
        m09\_Maoz & N/A & N/A & $4.2\times10^{4}$ & $11.39$ & $\approx0$ & $-3.14$ & $0.27$ & $7.2$ & $0.1$ & $0.33$ \\
        m09\_Z05-10x & $-0.5$ & $10\times$ & $1.9\times10^{4}$ & $12.71$ & $\approx0$ & $-3.01$ & $-0.37$ & $7.8$ & $0.2$ & $0.32$ \\
        m09\_Z05-100x & $-0.5$ & $100\times$ & $2.3\times10^{4}$ & $13.47$ & $\approx0$ & $-3.15$ & $-0.71$ & $10.9$ & $0.1$ & $0.27$ \\
		\hline
		\hline
\end{tabular}
\label{tab:modified-sims}
\begin{tablenotes}
\item $^\dagger$ Re-simulations with the dagger symbol use the \citet{mannucci-06} DTD; rest use the one from \citet{maoz-17}.
\end{tablenotes}
\end{table*}

\subsection{Observed Ia supernova rates}
\label{sec:2.5-ia-observations-all}

The two primary observational datasets of Ia rates we compare against throughout this paper are from ASAS-SN \citep{brown-19} and DES \citep{wiseman-21}. The quantity actually measured in both surveys is the number of $z\sim0$ Ia supernovae per unit time in galaxies in a given mass bin. They then divide these values by the total stellar mass in that bin from an independently measured galaxy stellar mass function (SMF) at $z\sim0$ to get a specific rate of Ia. This choice of assumed SMF is important, because it directly (linearly) affects the inferred specific rate of Ia.

\citet{brown-19} use the SMF from \citet{bell-03}, which is a fairly `shallow' mass function, with $dN/d\log M_*$ almost flat towards lower stellar masses. On the other hand, \citet{wiseman-21} use the \citet{baldry-12} SMF, which is much steeper towards the low-mass end -- roughly $dN/d\log M_* \sim M_*^{-1/2}$. This leads to a discrepancy in the total stellar mass in different mass bins assumed in both studies, especially at lower masses.
As we will show, \textit{this difference in assumed stellar mass function accounts for almost all of the apparent difference in specific Ia rate between ASAS-SN and DES in the regime $< 10^9\,\rm{M}_{\odot}$.}

Additionally, the choice of SMF influences the resultant abundance-matching relation in models like UM and requires different galaxy SFHs to be self-consistent. Specifically, a steeper SMF requires higher late-time SFRs to match the shallower $M_*$-$M_{\rm halo}$ relation, while choosing a shallower SMF requires a steeper $M_*$-$M_{\rm halo}$ relation and thus lower late-time SFRs, especially at lower masses. This also affects the resultant Ia rate, because supernova rates at $z\sim0$ are sensitive to recent star formation.

\begin{figure}
\centering
\begin{tabular}{c}
\includegraphics[width=1.0\linewidth]{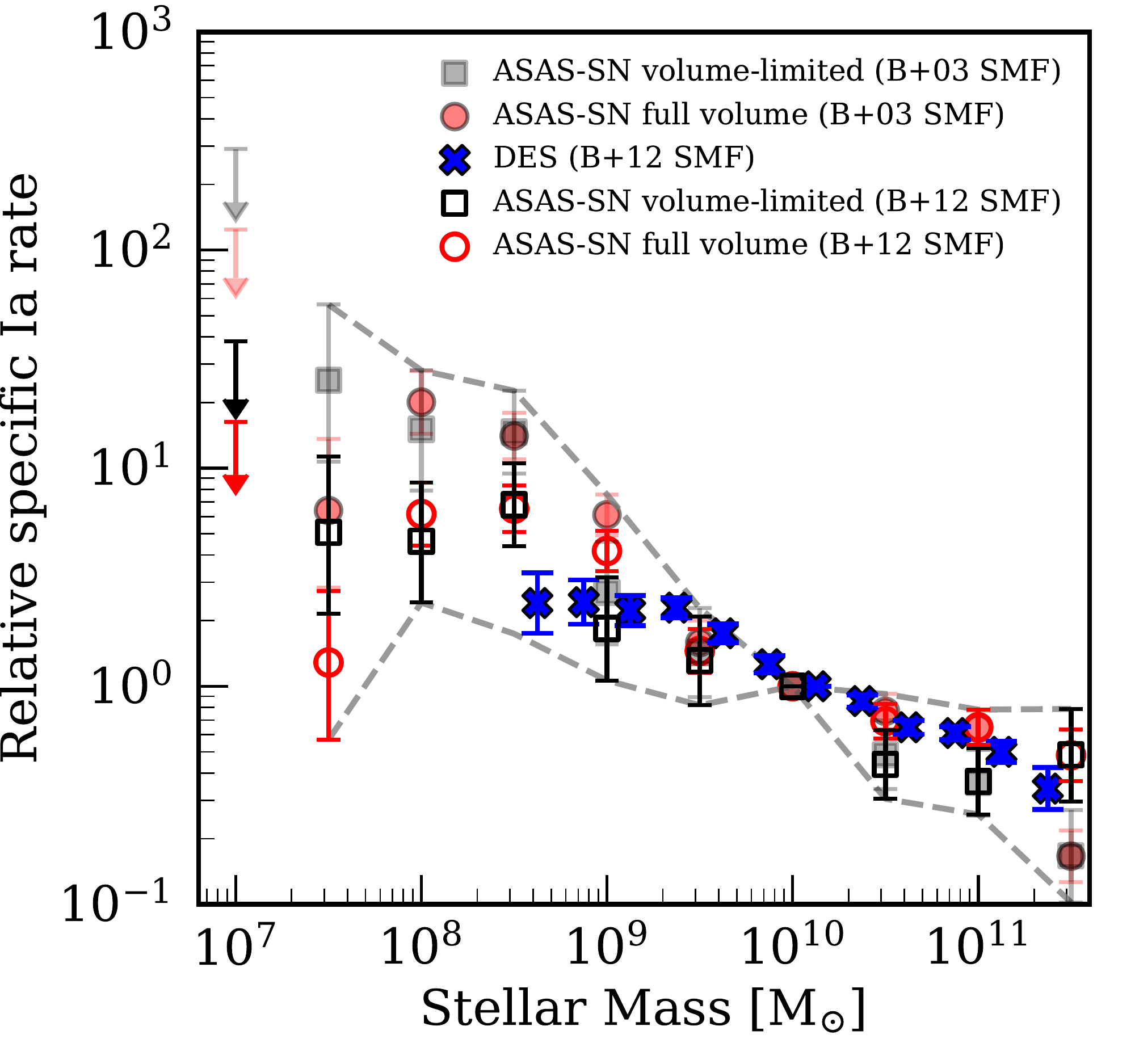}
\end{tabular}
\vspace{-2 mm}
\caption{
\textbf{Relative specific rates of Ia supernovae from different surveys.} In all cases we normalise to the same (arbitrary) rate at $M_* = 10^{10}\,\rm{M}_{\odot}$. Black filled squares and red filled circles show the original rates \citep[using the shallower SMF from][]{bell-03} from ASAS-SN, for their volume-limited (VL) and full-volume (FV) samples respectively \citep{brown-19}. The black unfilled squares and red unfilled circles show ASAS-SN rates modified to use the steeper SMF from \citet{baldry-12}, again for both the VL and FV samples. Finally, the blue points show the rates from DES which rely on the \citet{baldry-12} SMF. We demonstrate that the use of a steeper SMF for ASAS-SN values leads to much better agreement with DES rates, and a shallower trend versus stellar mass. The choice of SMF at $z\sim0$ strongly influences observed trends in Ia rates versus stellar mass. \textit{The grey dashed lines represent contours surrounding the full spread in all possible observed trends in Ia rate versus stellar mass (including systematics). In all subsequent figures, we show this full spread as a grey shaded region to compare our models against, along with the ASAS-SN volume-limited rates re-normalised to \citet{baldry-12} for reference.}
}
\label{fig:obs-rates}
\end{figure}

In Figure \ref{fig:obs-rates}, we compare specific rates of Ia supernovae from ASAS-SN and DES. We normalize all rates (simulated and observed) to the same (arbitrary) rate for galaxies at $10^{10} \rm{M}_{\odot}$. We do this because of the convention in \citet{brown-19}, as computing absolute Ia rates from observations is non-trivial, and our primary goal is to explore mass dependence of the rates, not their exact normalisation.

At masses $> 10^9\,\rm{M}_{\odot}$, the trend in specific rate from DES qualitatively agrees with the original ASAS-SN values (those that utilise the \citet{bell-03} SMF), for both their samples, rising by a factor of $2-6$ from $M_* \sim 10^{11}\,\rm{M}_{\odot}$ down to $\sim 10^9\,\rm{M}_{\odot}$. However, the two are discrepant for galaxies $< 10^9\,\rm{M}_{\odot}$, up to a factor of $7-8$. Both original ASAS-SN samples that utilise the shallower \citet{bell-03} SMF show significantly higher Ia rates at lower stellar masses compared to DES ($\sim10\times$ at $M_* ~ 3\times10^8 \rm{M}_{\odot}$, and larger at even lower masses). We also show modified versions of both ASAS-SN samples that use the \citet{baldry-12} SMF instead, which are more consistent with DES results. We compute this by multiplying the ASAS-SN rates in each mass bin by the ratio of the total stellar mass in that bin from the \citet{bell-03} SMF to that from the \citet{baldry-12} SMF, as Equation \ref{eq:smf-modifier} below shows:

\begin{equation}
    \; \; \; \; \; \; \; \frac{dN_{\rm Ia}}{dtdM_*}\bigg|_{\rm{modified}} \; \; = \; \; \frac{dN_{\rm Ia}}{dtdM_*}\bigg|_{\rm{ASAS-SN}} \; \; \times \; \; \frac{M_* (\rm{B03})}{M_* (\rm{B12})}
    \label{eq:smf-modifier}
\end{equation} \newline

\textit{This modification leads to better agreement (within $1\sigma$) between ASAS-SN and DES for Ia rates, with almost all the discrepancy at lower masses eliminated.} The question of the trend in specific Ia rate versus stellar mass then comes down to that of the appropriate SMF at $z\sim0$. A steeper SMF results in higher SFRs for low-mass galaxies at late times and specific Ia rates that have shallower dependence on stellar mass. Conversely, a shallower SMF leads to lower late-time SFRs and higher Ia rates at the low-mass end. While a more thorough examination of this issue is beyond the scope of this paper, it is important to note as a significant source of systematic uncertainty. \textit{Note: henceforth, we use the ASAS-SN volume-limited rates rescaled to the \citet{baldry-12} SMF as our `reference' observed rate to compare against, while also showing the full systematic scatter in rates across different observations and assumed SMFs.}


\section{Results}
\label{sec:3-results}

\subsection{Comparisons to observed Ia rates}
\label{sec:3.1-compare-ia}

The DTD from \citet{maoz-17} is an updated, more physically motivated model with stronger observational support than that of \citet{mannucci-06}. We also find that computing Ia rates in our simulations using either of the two DTDs results in qualitatively the same (approximately flat) trend versus galaxy mass. Therefore, for all subsequent analysis with metallicity-independent Ia rates, we apply only the \citet{maoz-17} DTD in post-processing.

\subsubsection{Metallicity-independent Ia rates}
\label{sec:3.1.1-nomet-rates}

\begin{figure}
\centering
\begin{tabular}{c}
\includegraphics[width =  1.0\linewidth]{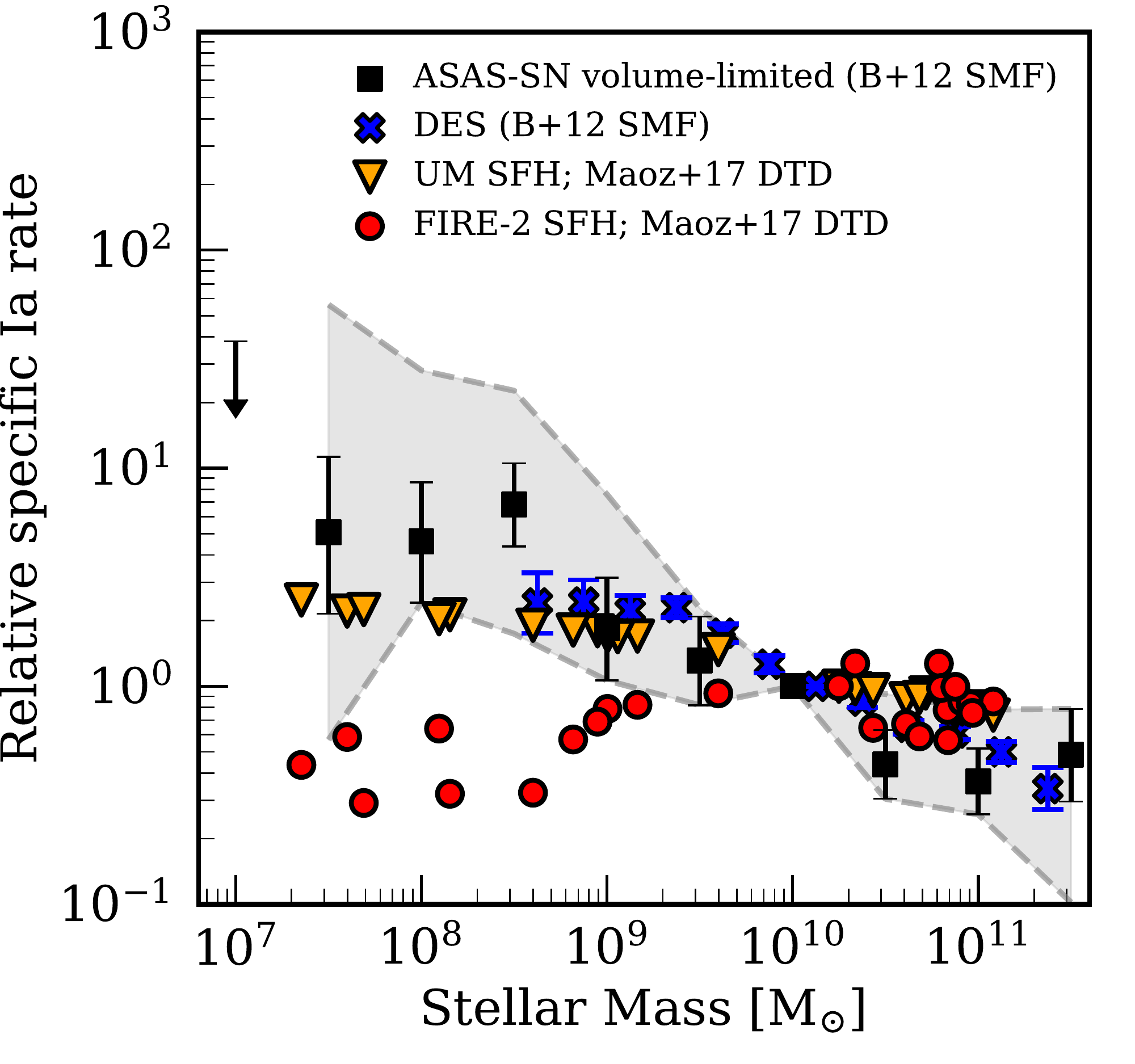}
\end{tabular}
\vspace{-2 mm}
\caption{
\textbf{Comparison of observed Ia specific rates to fidicial FIRE-2 simulations (with no metallicity dependence) and UM.}
We normalise all rates to the same (arbitrary) value at $M_* = 10^{10} \rm{M}_{\odot}$. Black squares show the rates from ASAS-SN using the \citet{baldry-12} SMF, as in Figure~\ref{fig:obs-rates}, with the grey shaded region showing the full spread in possible Ia rate trends. Blue points show rates from DES in \citet{wiseman-21}. Red circles show rates from FIRE-2 simulations, applying the DTD from \citet{maoz-17} in post-processing. These rates are discrepant with most of the possible observed trends across the entire mass range. Orange triangles show the rates by applying the same DTD to semi-empirical galaxy SFHs from \textsc{UniverseMachine} \citep[UM;][]{behroozi-19}. Going from $M_{\rm star}\sim10^{11}\rm{M_{\odot}}$ to $10^{7.5}\rm{M_{\odot}}$, the re-scaled ASAS-SN rates increase by $\sim20\times$, the fiducial FIRE-2 rates decrease by $\sim0.6$, and the UM rates increase by $\sim3\times$. DES values don't cover the entire mass range, but increase by $\sim4\times$ going from $10^{11}\rm{M_{\odot}}$ to $10^{8.5}\rm{M_{\odot}}$. \textit{While these semi-empirical SFHs help reduce the tension with observed rates somewhat, they still show a significant discrepancy compared to the stronger trends in the observations. This motivates our exploration of metallicity-dependent Ia rates.}
}
\label{fig:nomet-rates}
\end{figure}

Figure~\ref{fig:nomet-rates} compares observations of specific Ia rates to those from applying the metallicity-independent \citet{maoz-17} DTD to our simulations. For this figure and all subsequent ones showing Ia rates, we compare against the full systematic spread of allowed observational trends in specific Ia rate versus stellar mass from Figure~\ref{fig:obs-rates} (grey contours and shaded region), as well as the ASAS-SN values adjusted to the \citet{baldry-12} SMF as a representative sample. \textit{An important caveat about the grey shaded region is that just because it shows the full systematic spread in possible observations does not mean that one can simply fit the shallowest or steepest slope, and does not necessarily say that any single observational dataset is consistent with that entire range of slopes.}

As before, we normalise to the same (arbitrary) rate at $10^{10}\,\rm{M}_{\odot}$. Given that we benchmark sSFRs and $\tau_{90}$ ($90$ per cent stellar mass assembly timescale) values in our simulations at $z\sim0$ (Figures \ref{fig:SFMS-z0} and \ref{fig:tau-90}), and that our choices of DTD in FIRE-2 \citep{mannucci-06, maoz-17} are normalised to match observations, we know that the absolute Ia rates in our $M_* \sim 10^{10}\,\rm{M}_{\odot}$ galaxies are approximately consistent with observations, by design. Thus, we focus only on the relative trends in specific Ia rates here and in subsequent figures. 

Figure~\ref{fig:nomet-rates} shows a potential discrepancy between the simulated Ia rates using the fiducial DTD from \citet{maoz-17} and the rates from observations, especially with the stronger of the allowed trends. The potential discrepancy seems especially strong at low masses, with the FIRE-2 rates being $\approx 100\times$ lower than the upper end of the spread in observations at $\sim 10^7 \rm{M}_{\odot}$, and $\approx 10-20\times$ lower than ASAS-SN values using the \citet{baldry-12} SMF at the same mass. To check any possible variations (stochasticity) or systematic offsets in our simulated SFHs, we also show the rates computed by applying the same DTD to average galaxy SFHs \textsc{UM}. While the \textsc{UM} rates reduce the tension with observations somewhat, both the slope of the trend with mass and the relative rates still show a potential discrepancy at low masses. Going from $M_{\rm star}\sim10^{11}\rm{M_{\odot}}$ to $10^{7.5}\rm{M_{\odot}}$, the re-scaled ASAS-SN rates increase by $\sim20\times$, the fiducial FIRE-2 rates decrease by $\sim0.6$, and the UM rates increase by $\sim3-4\times$. DES values don't cover the entire mass range, but increase by $\sim4\times$ going from $10^{11}\rm{M_{\odot}}$ to $10^{8.5}\rm{M_{\odot}}$. Considering the full systematic spread in possible observed trends either somewhat decreases the discrepancy with the \citet{maoz-17} DTD or greatly increases it, given the logarithmic scale on the y-axis. \textit{The fact that these discrepancies exist regardless of the choice of SFHs further motivates our exploration of modified (metallicity-dependent) rates for Ia supernovae.}

\subsubsection{Metallicity-dependent Ia rates}
\label{sec:3.1.2-met-rates}

\begin{figure*}
\centering
\begin{tabular}{c c}
\includegraphics[width=0.48\linewidth]{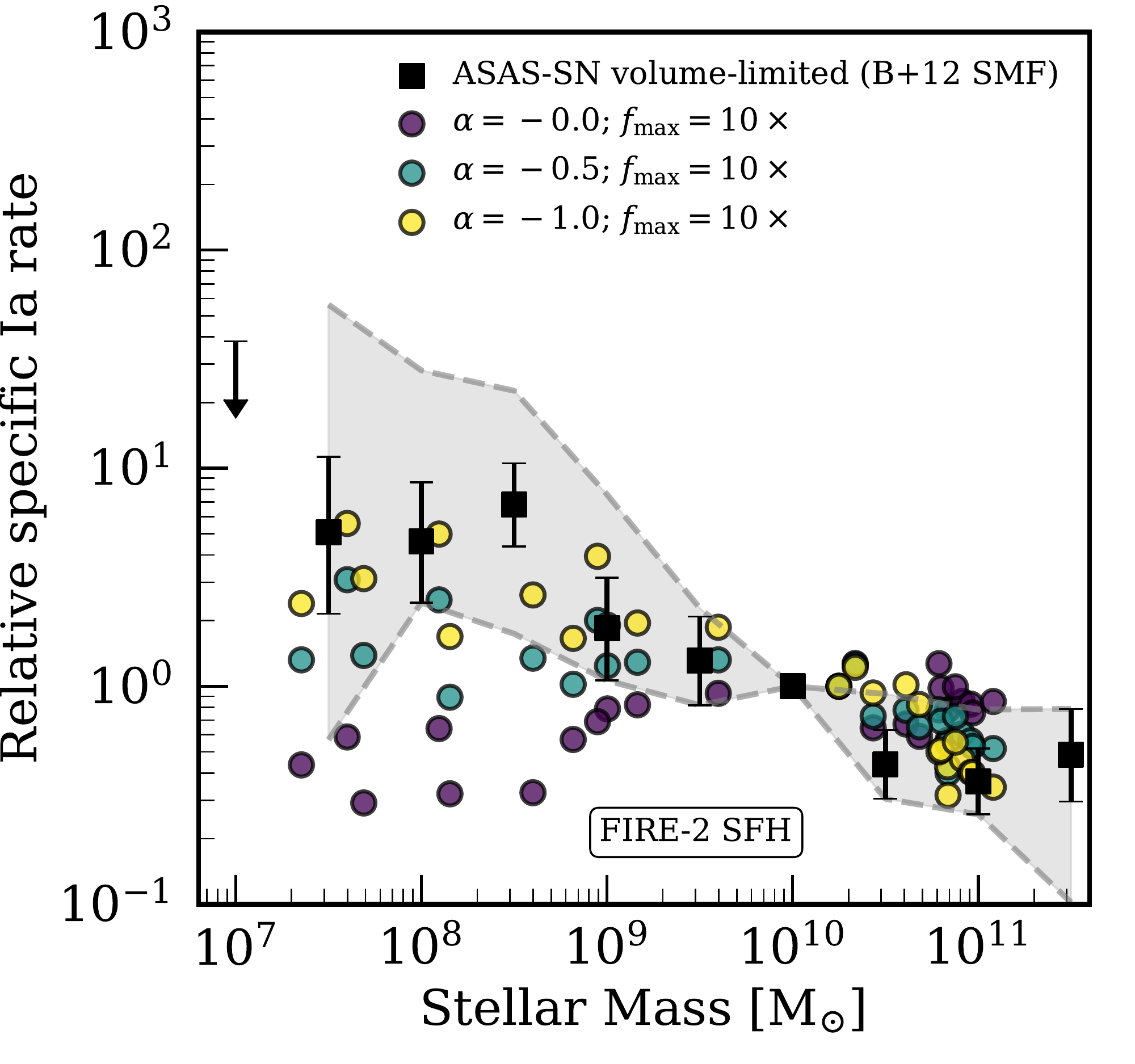}&
\includegraphics[width=0.48\linewidth]{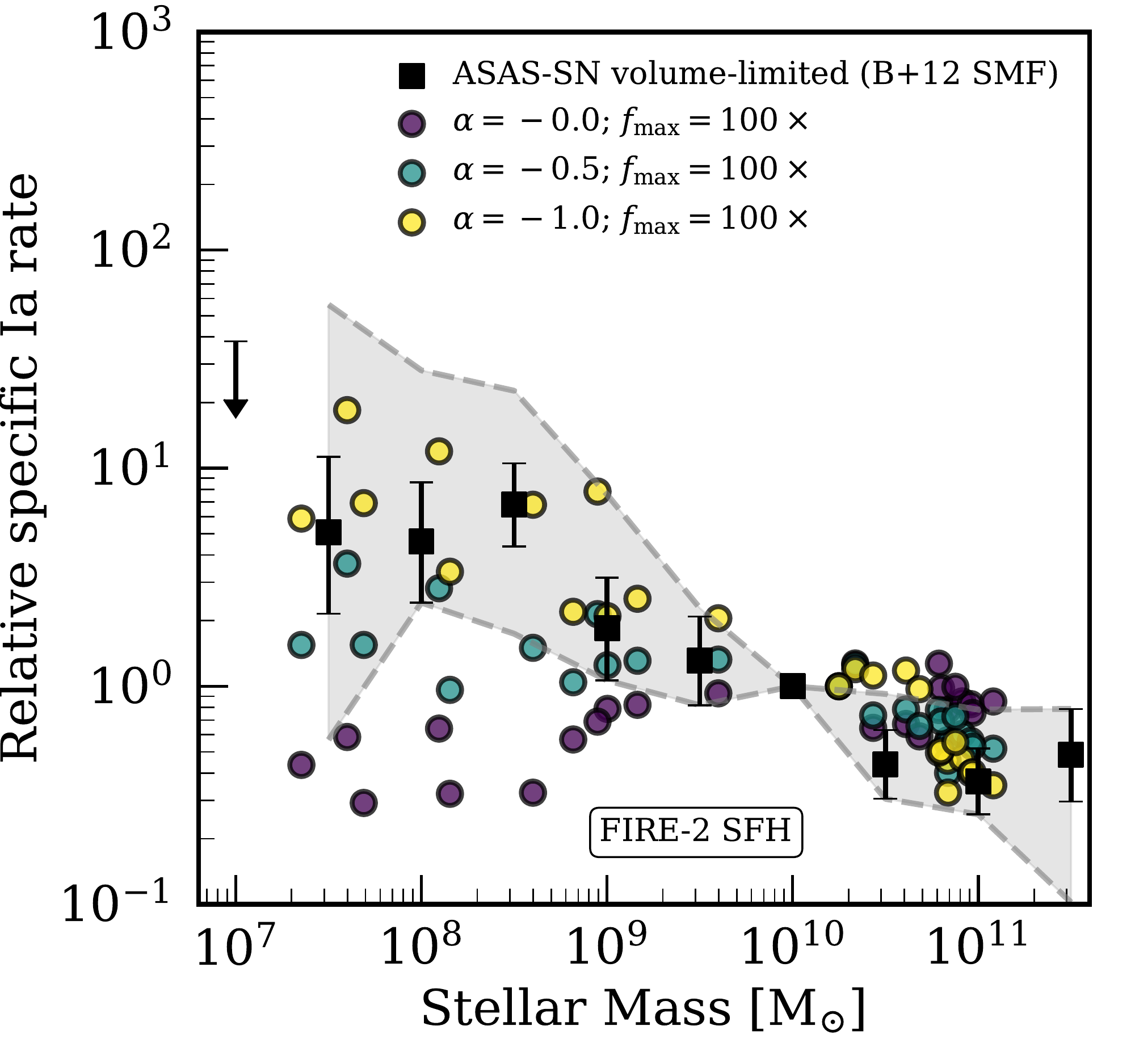} \\
\includegraphics[width=0.48\linewidth]{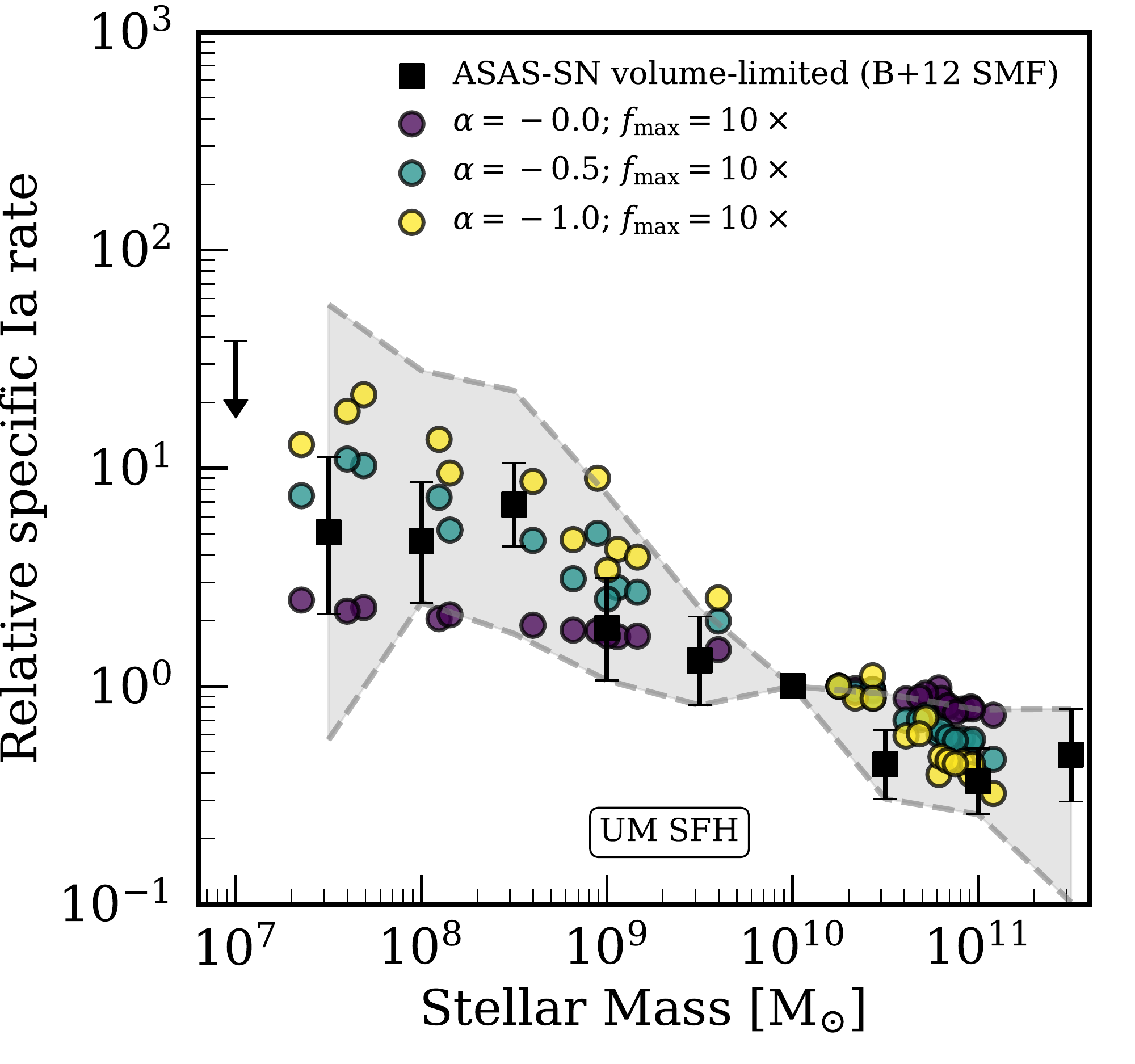}&
\includegraphics[width=0.48\linewidth]{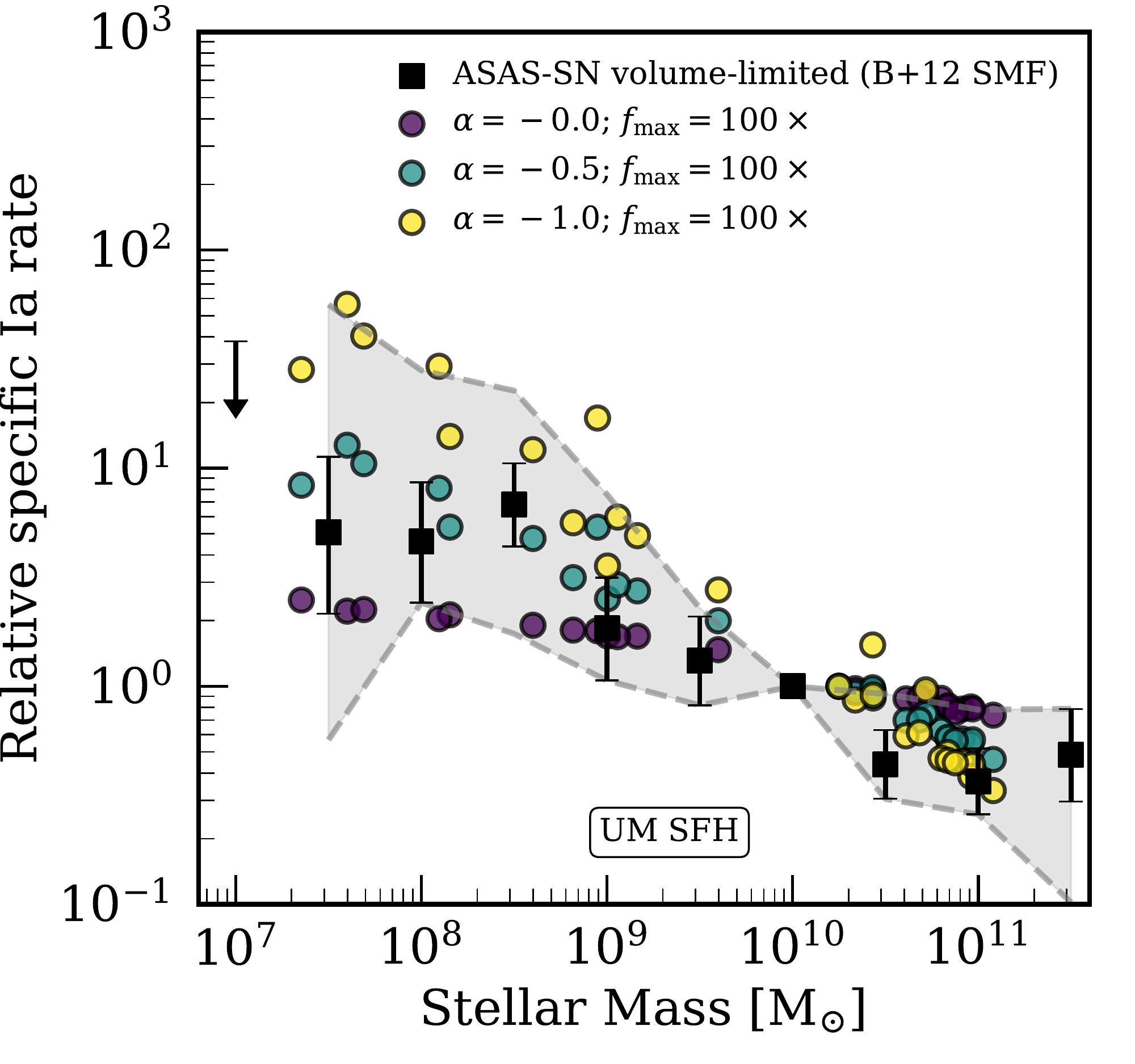}
\end{tabular}
\caption{
\textbf{Metallicity-dependent modifications to the rates of Ia supernova}.
The results of testing power-law modifications to the normalisation of the \citet{maoz-17} DTD (as defined in section~\ref{sec:2.4-ia-rates-met}). $\alpha$ is the power-law exponent, and the rate boost cap ($f_{\rm max}$) ensures that no population has its rate boosted by more than $10\times$ or $100\times$. We apply these modified Ia rates in post-processing, assuming that they do not significantly affect the SFHs (but see also Figure~\ref{fig:newrun-rates}).
As in Figure~\ref{fig:nomet-rates}, black squares show the rates from ASAS-SN using the \citet{baldry-12} SMF, with the grey shaded region showing the full spread in possible Ia rate trends versus stellar mass.
\textbf{Top left}: Simulated rates with FIRE-2 SFHs and $f_{\rm max}=100\times$.
\textbf{Top right}: Same as top left, but with $f_{\rm max}=10x$. \textbf{Bottom left}: Same, but re-normalizing the star particle ages to match the mean SFH from \textsc{UM}, thus preserving the metallicity distributions in each simulation, and with $f_{\rm max}=10x$.
\textbf{Bottom right}: Same as bottom left, but with $f_{\rm max}=100x$.
With FIRE-2 SFHs, we require strong metallicity dependence ($\alpha \sim -1.0$, $f_{\rm max}=100 \times$) to match the strongest observed trends, while modifying the SFHs to match \textsc{UM} requires less extreme dependence on metallicity ($\alpha=-0.5\,\rm{to}\,-1.0$, $f_{\rm max}$ as low as $10 \times$).
\textit{Thus we find a potentially viable parameter space, in agreement with observations, for metallicity-dependent rates of Ia supernovae.}
}
\label{fig:met-rates}
\end{figure*}


Figure~\ref{fig:met-rates} shows our metallicity-dependent modifier to Ia rates applied to the FIRE-2 simulations \textit{in post-processing} (we test self-consistent re-simulations in Section~\ref{sec:3.2.1-resim-ia-rates}), compared to the same range in allowed observational trends as in Figure~\ref{fig:nomet-rates}. For our models, we show Ia rates based on power-law exponent values (see Equation~\ref{eq:met-modifier}) of $\alpha=0.0$, $\alpha=-0.5$, and $\alpha=-1.0$. The top two panels of Figure~\ref{fig:met-rates} show metallicity-dependent rates using the fiducial FIRE-2 SFHs, along with a rate boost cap of $f_{\rm max} = 10 \times$ or $100\times$. Reproducing the strongest allowed trend in the observations requires a fairly extreme metallicity dependence of $\alpha \approx -1.0$ and $f_{\rm max} \gtrsim 100\times$. 

The bottom two panels of Figure~\ref{fig:met-rates} show the application of the metallicity-dependent modifiers to Ia rates to SFHs from UM instead. We do this by retaining the metallicities of the star particles in FIRE-2 galaxies but adjusting their ages to match the SFHs of the UM galaxy closest in stellar mass at  $z=0$. While this is not fully self-consistent, given that the elemental enrichment history of a galaxy may change with its SFH, it does preserve the relative rank ordering of age and metallicity among star particles, and we consider it a reasonable test given the absence of a rigorous semi-empirical model for enrichment. With these SFHs (that have more recent star formation in lower $\rm{M}_*$ galaxies, hence a more enhanced Ia rate there), a power-law exponent of $\alpha = -0.5$ is sufficient to match ASAS-SN values using the \citet{baldry-12} SMF, and that of $\sim -1.0$ is sufficient to reproduce the strongest observed trend in Ia rate, depending on the choice of $f_{\rm max}$.

\begin{figure}
\centering
\begin{tabular}{c}
\includegraphics[width =  0.98\linewidth]{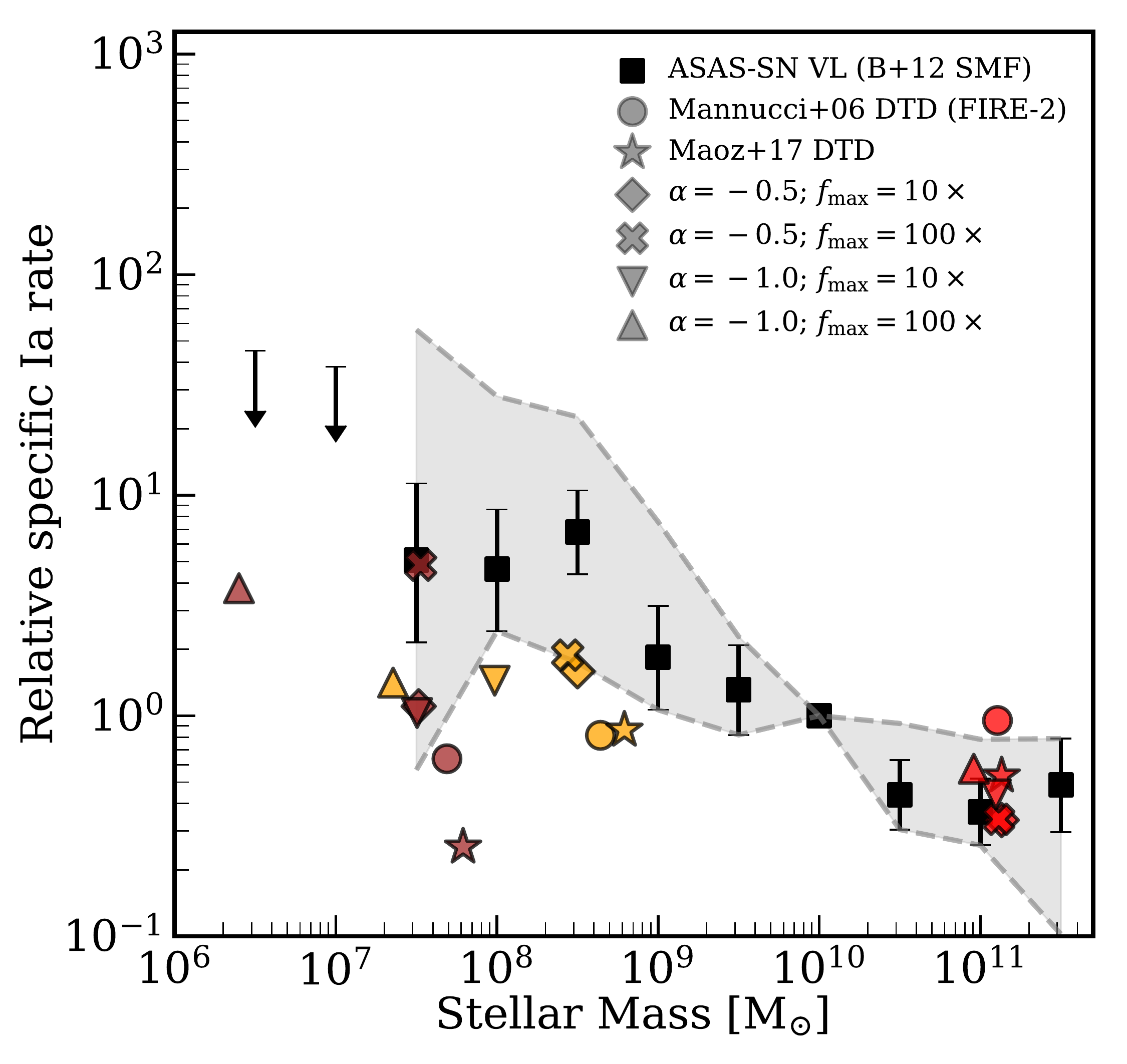}
\end{tabular}
\vspace{-2 mm}
\caption{
\textbf{Specific rates of Ia supernovae for FIRE-2 re-simulations with different DTD models.}
Black squares show the rates from ASAS-SN using the \citet{baldry-12} SMF, with the grey shaded region showing the full spread in possible Ia rate trends versus stellar mass. We show 3 of the re-simulated \textsc{FIRE-2} galaxies with different metallicity-dependent models for Ia rates as described in Table~\ref{tab:modified-sims}: m12i (red points), m11e (orange points), and m11b (brown points). Circles and stars show metallicity-independent \citet{mannucci-06} and \citet{maoz-17} models, while the other shapes show the metallicity-dependent models. The re-simulations with stronger metallicity dependence to Ia rates show better agreement with the observed trends, though the lower-mass galaxies re-simulated with the most extreme rate model ($\alpha=-1.0$; $f_{\rm max}=100 \times$) form lower stellar mass by up to an order of magnitude, given the significantly increased stellar feedback. \textit{Thus, self-consistent feedback limits how steep of a relation between Ia rate and stellar mass one can achieve, even as $\alpha << 0$ and $f_{\rm max} \rightarrow \infty$. However, this comparison does allow us to favour our moderate model ($\alpha=-0.5$, $f_{\rm max}=100\times$) since it gives us rates within observational scatter while not impacting the final stellar mass to a large extent.}
}
\label{fig:newrun-rates}
\end{figure}

\subsection{Metallicity-dependent models for Ia rates in self-consistent galaxy re-simulations}
\label{sec:3.2-impact-on-galaxy-formation}

\subsubsection{Self-consistent metallicity-dependent Ia rates}
\label{sec:3.2.1-resim-ia-rates}

Having shown that metallicity-dependent Ia rate modifiers are able, in principle, to produce simulated rates that are consistent with the full range allowed by observations, we seek to examine the effects of boosting Ia rates on other aspects of galaxy formation. Re-simulating galaxies with different Ia rate models is fundamentally different from simply computing rates in post-processing. In the former, changes to the Ia rate affect both mechanical feedback and metal enrichment, leading to differences in SFHs as well as enrichment histories, while in the latter (and all of our analysis so far) we simply compute Ia rates without altering galaxy properties.

Self-consistent re-simulations are especially important for lower-mass galaxies, with more metal-poor stellar populations, which would have their rates boosted much more than the Milky-Way mass galaxies. As explained in Section~\ref{sec:2.4-ia-rates-met}, we re-simulate 4 of the FIRE-2 galaxies (m12i, m11e, m11b, and m09) with various metallicity-dependent modifiers to the \citet{maoz-17} DTD. Table~\ref{tab:modified-sims} lists these re-simulations, with their Ia rate modifiers, and resultant galaxy properties.

Figure~\ref{fig:newrun-rates} shows the specific Ia rates for the re-simulations of 3 of the galaxies run with our metallicity-dependent rate models. Importantly, the effect on the rate versus $M_*$ diagram is not quite the same as that from our post-processing analysis in Figure~\ref{fig:met-rates}, especially for the lower-mass galaxies (m11e and m11b), because the re-simulations with extreme rate models ($\alpha=-1.0$ and $f_{\rm max}=100\times$) form significantly fewer stars, owing to increased supernova feedback. This means that the rates in the re-simulated galaxies are plausibly consistent with the shallowest observed trends and quasi-consistent with the re-scaled ASAS-SN rates, but not with the steepest possible observed trends. However, they allow us to favour our moderately metallicity-dependent model ($\alpha=-0.5$, $f_{\rm max}=100\times$) since it gives us rates within observational scatter while not having too large of an impact on the final galaxy masses. 


\subsubsection{Effects on galaxy properties: mass, size, morphology}
\label{sec:3.2.2-morphology}

We next examine the effect of our modified models for Ia rates on the overall properties of these galaxies.
For the re-simulations of m12i, m11e, and m11b, we compare 4 parameters at $z=0$: (a) stellar mass, (b) $R_{90}$, the radius enclosing $90$ per cent of the stellar mass, (c) $v / \sigma$ for stars, a measure of rotational versus dispersion support, and (d) minor to major axis ratio of the stars (based on the principal axes of the rotational inertia tensor of the galaxy). Note that the m09 re-simulations are at a mass scale especially sensitive to even the slightest perturbations in feedback energetics and with morphologies and sizes that are not well constrained by observations, so we do not include those in this part of the analysis.

\begin{figure}
\centering
\begin{tabular}{c}
\includegraphics[width=0.6\linewidth]{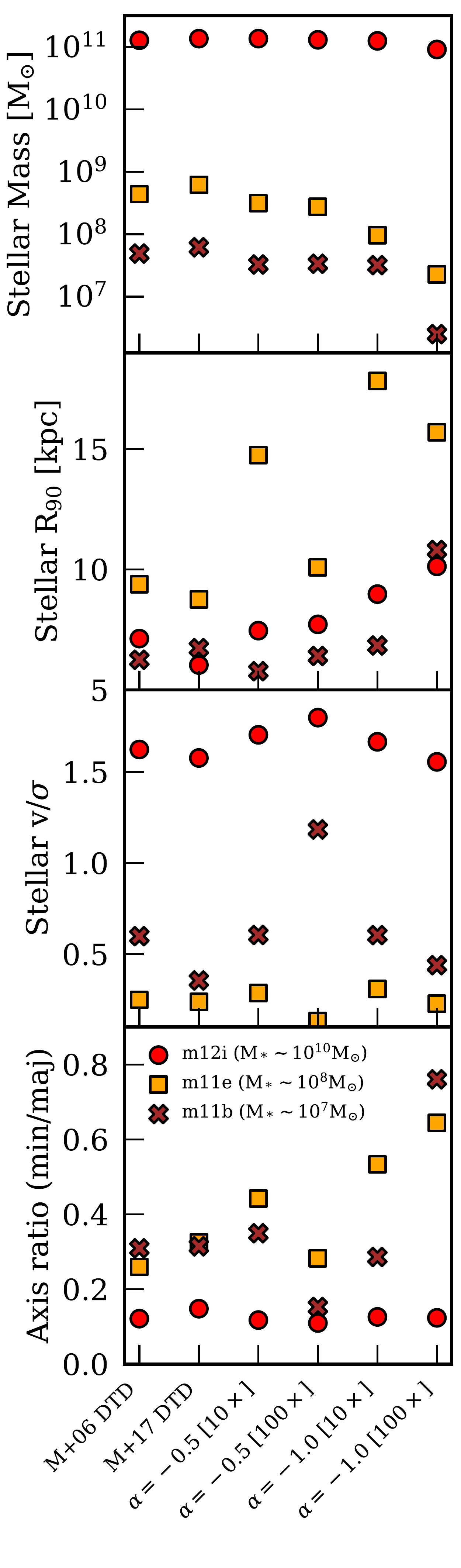}
\end{tabular}
\caption{
\textbf{Properties at $z=0$ in our metallicity-dependent re-simulations.}
We show the re-simulations of m12i (red circles), m11e (orange squares) and m11b (brown crosses).
\textbf{Stellar mass}: m12i variants show no appreciable change in stellar mass, because of the relatively minor boost in Ia rate at this metallicity scale. For m11e and m11b, the most extreme metallicity-dependent rate models cause a significant drop in stellar mass, by almost an order of magnitude.
\textbf{Sizes}: m12i and m11b show gradual increases in their radius enclosing $90$ per cent of stars with the more extreme rate models, while m11e shows a much more drastic change in $R_{90}$ and results in a more diffuse galaxy for the more extreme rate models.
\textbf{Rotation}: None of the 3 re-simulations show any appreciable change in their stellar $v / \sigma$ (`diskiness') for different rate models.
\textbf{Axis ratios}: m12i keeps its small axis ratio (more disk-like) in stars for all re-simulations, while m11e and m11b show gradual increases in their axis ratios: a trend towards more spheroidal galaxies with more extreme rate models.
}
\label{fig:newrun-morphology}
\end{figure}

Figure~\ref{fig:newrun-morphology} (top) shows that for the Milky Way-mass galaxy m12i, the various re-simulations do not significantly impact its stellar mass, because the Ia rates near solar metallicity scale are barely changed (by design).
Thus, any changes to its early formation (when it was low metallicity) do not significantly alter its properties at $z \sim 0$.
The lower-mass galaxies m11e and m11b, however, show a gradual decline in stellar mass with increasing Ia rate boosts, with the most extreme models leading to an almost order of magnitude drop. Figure~\ref{fig:newrun-morphology} (upper middle) shows $R_{90}$, which gradually increases with more extreme rate models for m12i and m11b. m11e shows a sharper increase in radius, with the most extreme metallicity-dependent models resulting in diffuse galaxies. We suspect that this is because some feature of m11e's formation history makes its size especially susceptible to changes in feedback energetics. For more detailed studies on `puffiness' of low-mass galaxies using the FIRE simulations, refer to \citet{el-badry-16} and \citet{kado-fong-21}. Figure~\ref{fig:newrun-morphology} (lower middle) shows that the stellar $v / \sigma$, or the rotation versus dispersion metric, remains nearly unchanged for all 3 galaxies regardless of the choice of model for Ia rates. Finally, Figure~\ref{fig:newrun-morphology} (bottom) shows that m12i remains `disk-like' (small axis ratios) for all its variant re-simulations, while m11e and m11b show increasing axis ratios and more spheroidal morphologies for the more extreme rate models.

We conclude that our most extreme metallicity-dependent Ia rate models most clearly impact the stellar masses and effective radii of the re-simulated galaxies, especially at lower-masses. However, there remains room for our more modest rate models that don't have as large of an impact on the resultant galaxy overall. For these properties, rigorously comparing against observations is beyond the scope of this paper -- but is potentially important for future work.

\subsubsection{Relation between stellar mass and metallicity}
\label{sec:3.2.3-MZR}

\begin{figure*}
\centering
\begin{tabular}{c c}
\includegraphics[width = 0.48 \linewidth]{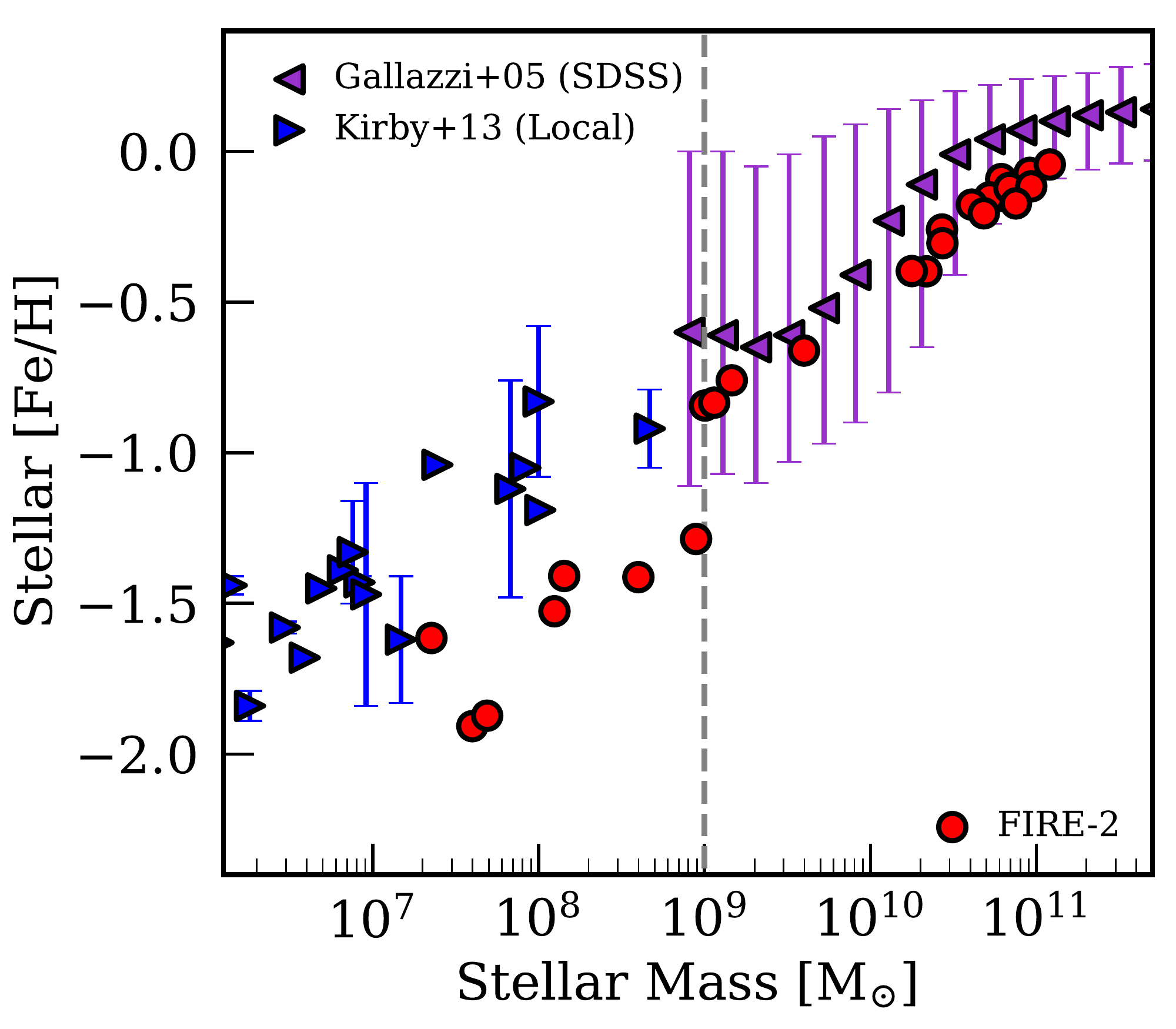}&
\includegraphics[width = 0.48 \linewidth]{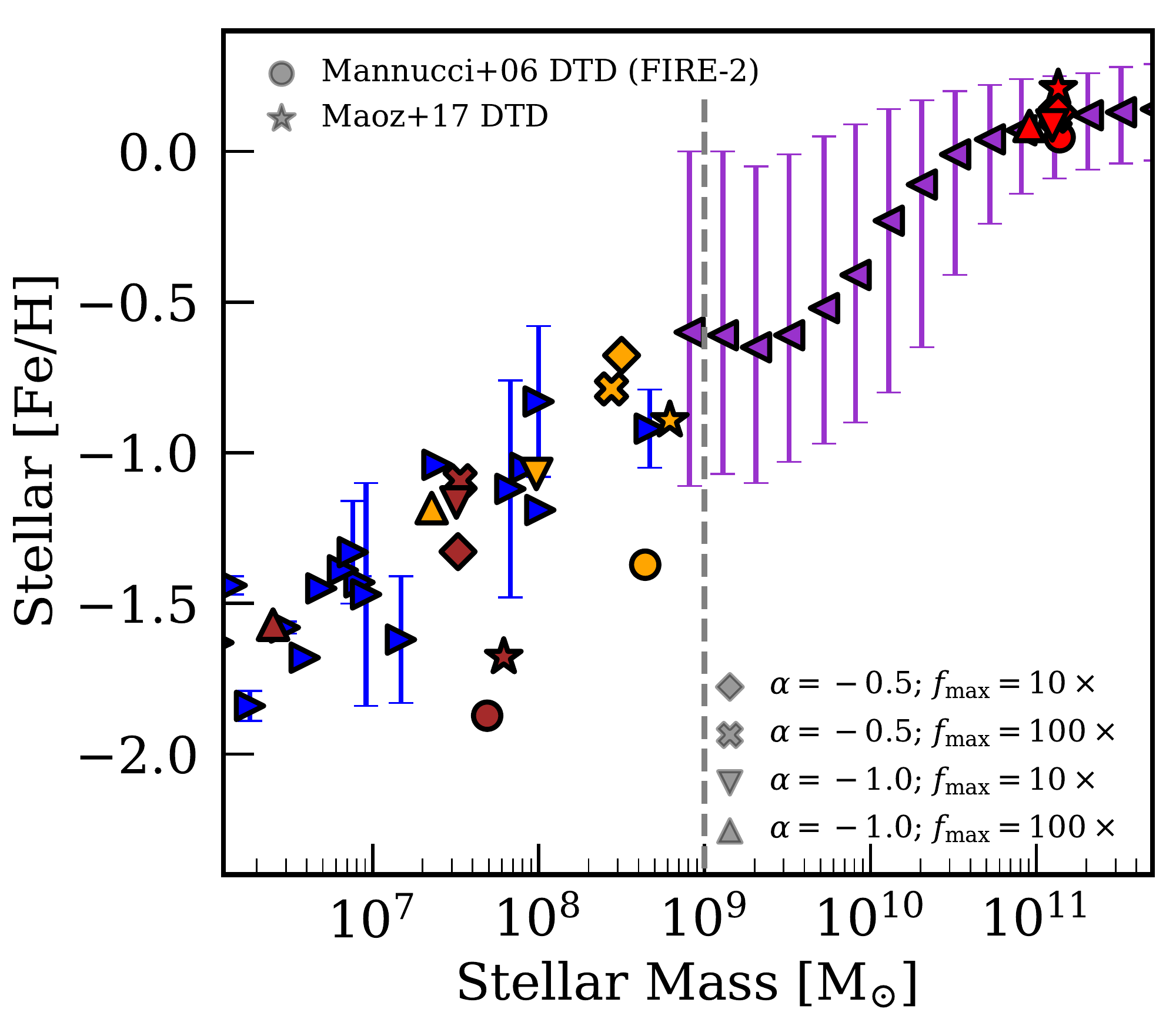}
\end{tabular}
\vspace{-2 mm}
\caption{
\textbf{Relation between stellar mass and stellar metallicity}.
We show observed values of [Fe/H] from \citet{gallazzi-05} and \citet{kirby-13} for reference. In both panels, values for simulations with $M_*>10^9\,\rm{M}_{\odot}$ are $M_*$-weighted linear mean metallicities, while those for simulations with $M_*<10^9\,\rm{M}_{\odot}$ are $M_*$-weighted median values (as described in Section~\ref{sec:3.2.3-MZR}). Vertical grey dashed lines partition the two regimes.
\textbf{Left}: mass-metallicity relation for our fiducial \textsc{FIRE-2} suite (red circles). \textsc{FIRE-2} galaxies show good agreement with observations $M_* \geq 10^{9} \rm{M}_{\odot}$, but $> 1\sigma$ discrepancy for $M_* < 10^{9} \rm{M}_{\odot}$, on average.
\textbf{Right}: same, for 3 of the galaxies that we re-simulate with various metallicity-dependent models for Ia rates: m12i in red, m11e in orange, and m11b in brown. We show the original \textsc{FIRE-2} galaxies with the \citet{mannucci-06} DTD as circles, the re-simulations with the metallicity-independent DTD from \citet{maoz-17} as squares, and those with the metallicity-dependent Ia rates as other shapes. 
\textit{For $M_* \leq 10^9\,\rm{M}_{\odot}$, re-simulations with metallicity-dependent Ia rates show better agreement with observations.}
}
\label{fig:MZR-massive}
\end{figure*}

We also examine the impact on stellar elemental abundances.
Figure~\ref{fig:MZR-massive} (left) shows the present-day relation between stellar metallicity, [Fe/H], and stellar mass for our entire suite of original FIRE-2 galaxies. We compare to observed stellar iron abundances from \citet{kirby-13} at low mass and \citet{gallazzi-05} at high mass. While \citet{gallazzi-05} measured metallicities of high-mass galaxies with SDSS fiber spectroscopy, \citet{kirby-13} measured resolved star-by-star metallicities in Local Group low-mass galaxies. Therefore, to make accurate comparisons, we compute iron abundances for our simulated galaxies in 2 ways: (a) using the logarithmic value of $M_*$-weighted linear mean metallicity for galaxies with $M_*>10^9\,\rm{M}_{\odot}$, and (b) a median value of log iron abundances of individual star particles, weighted by the particle masses. Equation~\ref{eq:met-SMWLM} describes the log $M_*$-weighted linear mean metallicity, where the sums are over star particles (denoted by $i$), ${Z_{\rm Fe}}$ represents the iron mass fraction of star particles, and $1.38\times10^{-3}$ is the Solar iron mass fraction from \citet{asplund-09}.

\begin{equation}
    \; \; \; \; \; \; \; \; \; \; \rm{Metallicity} \;=\; \log_{10} \left[ \frac{\sum_i \left(M_{\rm{star,i}} \times Z_{\rm{Fe,i}}\right)}{\sum_i \left(M_{\rm{star,i}} \times 1.38\times10^{-3}\right)} \right]
    \label{eq:met-SMWLM}
\end{equation}\newline 

We find that iron abundances for FIRE-2 galaxies at stellar masses $>10^9\,\rm{M}_{\odot}$ agree with observations within $1\sigma$, while galaxies at lower masses show a systematic deficit relative to the observed relation. Previous studies of FIRE-2 simulations have shown this offset at $M_* < 10^9 \,\rm{M}_{\odot}$ as well, such as \citet{ma-16}, \citet{wetzel-16}, \citet{escala-18}, \citet{wheeler-19}, and \citet{hopkins-20}.

The right panel of Figure~\ref{fig:MZR-massive} shows the same stellar mass-metallicity relation, but for our re-simulations of m12i, m11e, and m11b with metallicity-dependent rates. The metallicities of the m12i re-simulations hardly change, because the boost in Ia rates at this mass is negligible -- the good agreement with observations remains. For the lower-mass m11e and m11b re-simulations, the switch from the \citet{mannucci-06} DTD to the metallicity-independent DTD from \citet{maoz-17} leads to a modest but significant increase in [Fe/H], because of the net increase in the total number of Ia supernovae. A previous study using FIRE simulations \citep{muley-21} also reported this increase in total number by simply switching to the \citet{maoz-17} DTD. Additionally, the metallicity-dependent rate models lead to improved agreement with the observed mass-metallicity relation. For the most extreme rate models ($\alpha=-1.0$ with $f_{\rm max}=10\times$ or $100\times$), the overall drop in stellar mass leads to the galaxies moving down along the relation, while still agreeing well with the observed values. This suggests that the systematic discrepancy in low-mass, low-metallicity galaxies is possibly from underestimation of Type Ia supernova rates, and our intermediately metallicity-dependent models can account for any apparent discrepancy at these masses.


\subsubsection{Iron abundances in ultra-faint galaxies}
\label{sec:3.2.4-FeH-UFD}

\begin{figure}
\centering
\begin{tabular}{c}
\includegraphics[width = 1.0 \linewidth]{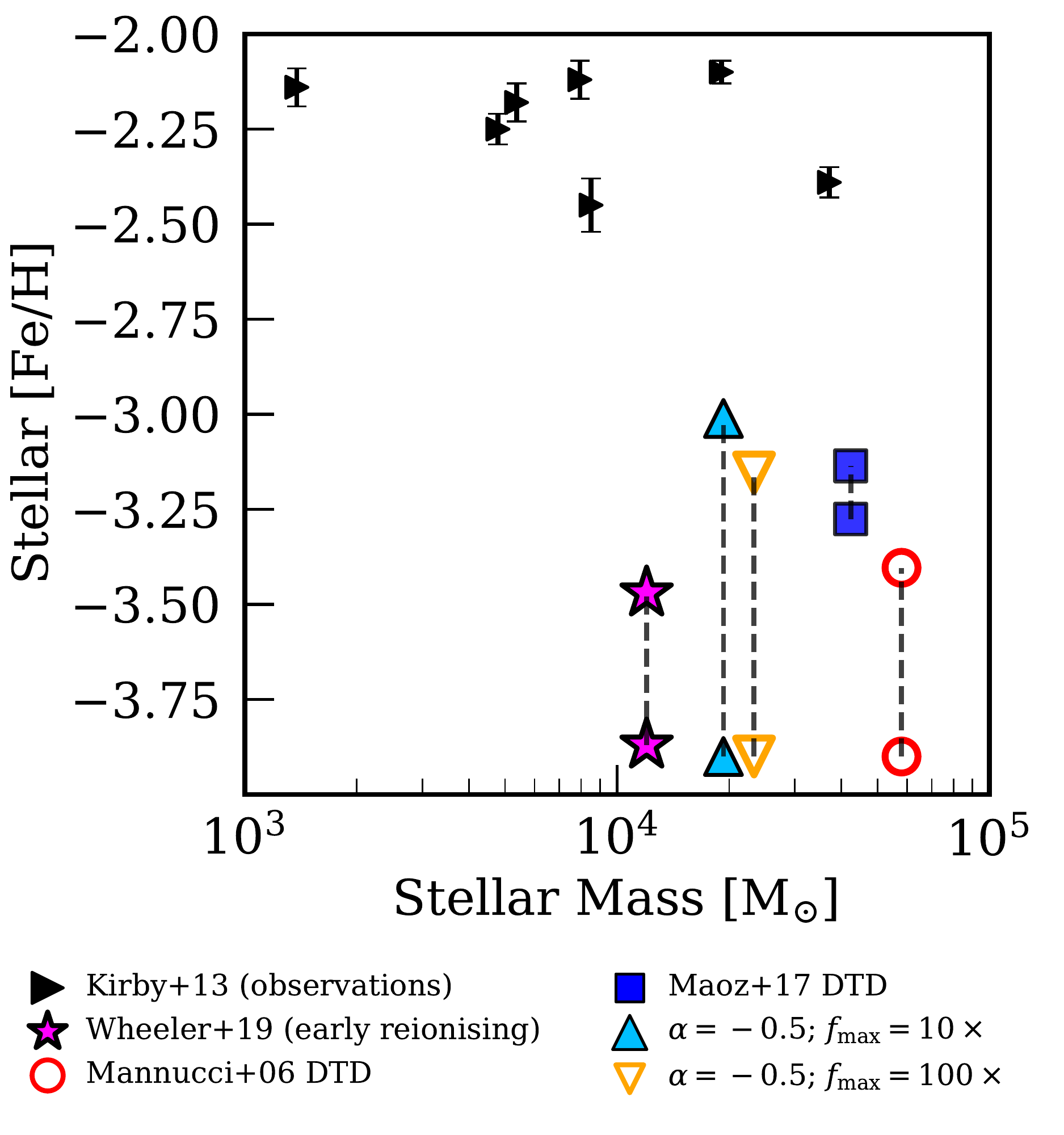}
\end{tabular}
\vspace{-2 mm}
\caption{
\textbf{Relation between stellar mass and stellar metallicity for ultra-faint galaxies}.
Black triangles show observations compiled in \citet{kirby-13}, while other points show original and re-simulations of m09 as Table~\ref{tab:modified-sims} lists. For each re-simulation, we show 2 values for the median iron abundance, connected by dashed grey lines in each case - the lower values include star particles that form at our initial (artificial) metallicity floor of [Fe/H]$= -3.82$, while the upper values exclude these star particles. The pink stars show the original FIRE-2 m09 presented in \citet{wheeler-19}, which uses the \citet{mannucci-06} DTD and the early-reionising UV background from \citet{faucher-giguere-09}, with $z_{\rm reion} \sim 10$. Red circles show m09 re-simulated with an updated, later-reionising UV background \citep{faucher-giguere-20}, with $z_{\rm reion} \sim 7.8$. Blue squares show a re-simulation using the updated UV background and a metallicity-independent DTD from \citep{maoz-17}. Finally, the blue triangles and orange inverted triangles show our re-simulations with metallicity-dependent DTDs. Unfilled points are re-simulations at lower resolution ($M_{\rm{baryon}} = 250 \rm{M}_{\odot}$), while filled ones are at full resolution of $M_{\rm{baryon}} = 30 \rm{M}_{\odot}$. \textit{The re-simulations with metallicity-dependent models for Ia rates improve the agreement with observations relative to that of \citet{wheeler-19}, especially if we only consider only enriched stars.}
}
\label{fig:MZR-UFD}
\end{figure}

At stellar masses of $10^3 - 10^5 \, \rm{M_{\odot}}$, previous studies \citep[][for example]{wheeler-19}, have shown that simulated galaxies tend to significantly under-produce their iron abundance relative to observations such as those from \citet{kirby-13}. Simulation projects besides FIRE also have shown this under-enrichment of ultra-faint galaxies by $\sim 2$ dex or more relative to \citet{kirby-13}, such as \citet{maccio-17}, the Engineering Dwarfs at Galaxy formation's Edge \citep[EDGE;][]{agertz-20} project, and the Mint Resolution DC Justice League simulations \citep[][]{applebaum-21}. We investigate whether introducing metallicity-dependent models for Ia rates alleviates this tension, using the re-simulated versions of m09.

Figure~\ref{fig:MZR-UFD} shows the relation between stellar [Fe/H] and stellar mass at $10^3 - 10^5 \, \rm{M_{\odot}}$. We also show observations in \citet{kirby-13}. For comparison to prior studies using the FIRE-2 simulations, we include the point from \citet{wheeler-19} for the original m09 simulation, which used the standard cosmic UV background in FIRE-2 \citep[][]{faucher-giguere-09}, corresponding to an earlier redshift of HI reionisation of $z_{\rm reion}\sim10$. We also show median [Fe/H] abundances for our version of the m09 simulation with an updated cosmic UV background model \citep[][]{faucher-giguere-20}, corresponding to a later reionisation redshift of $z_{\rm reion}\sim7.8$, as well as the versions of it re-simulated with different metallicity-dependent models for Ia rates.

For all simulations, we show 2 different $M_*$-weighted median [Fe/H] values, connected by a grey dashed line in each case. This is because the FIRE-2 model does not model Pop III stars, instead assigning a metallicity floor of [Fe/H] $= -3.82$ to pristine gas when simulations begin. (This value is based on \citet{asplund-09} Solar values, which corresponds to $-4.0$ in units of \citet{anders-89}.) At these extremely low masses, reionisation plus stellar feedback result in short SFHs with few enriched stars forming, such that many star particles form from gas at the metallicity floor, which results in the median [Fe/H] value being artificially skewed towards $\sim -4$. Because we cannot adjust for this in the absence of a full model of Pop III stars, and because Pop III stars would not survive to $z = 0$ to have their [Fe/H] measured, we show two median [Fe/H] values for each galaxy - one that takes into account all star particles and another that only considers star particles that have been enriched above the metallicity floor.

A couple of other considerations also potentially justify considering the enriched populations: (a) even at the full $30\, \rm{M}_{\odot}$ resolution, our simulations may not be fully time-resolving the SFHs of these extremely low-mass galaxies to resolve early enrichment events, and (b) uncertainties surrounding variations in early stellar initial mass functions (IMFs) suggest that the low-mass end may not be fully populated at low metallicities.


Note that our modifications to the UV/X-ray ionising background and Ia rates in m09 have a non-trivial impact on its $z=0$ stellar mass -- these impacts, however, make sense given the underlying changes in physics. The first-order effect is that of the updated ionising background; simply changing the reionisation timeline from $z_{\rm reion} \sim 10$ to the updated one with $z_{\rm reion} \sim 7.8$, that is, going from the pink \citet{wheeler-19} point to the unfilled red \citet{mannucci-06} point leads to an increase in $M_{\rm star}$ by a factor of $5 - 6\times$. This makes sense because the impact of the reionisation timeline on star formation is largest at this mass scale. Going from the unfilled red to the filled blue point leads to a small decrease in $M_{\rm star}$ because of the modest increase in overall numbers of Ia supernovae. Finally, the significant boosts to the Ia rates in the metallicity-dependent model re-simulations lead to further decrease in $M_{\rm star}$, and the $f_{\rm max}=10\times$ and $f_{\rm max}=100\times$ versions are similar to each other within run-to-run numerical stochasticity. An interesting coincidence is that adding metallicity dependence to Ia rates leads to a final $M_{\rm star}$ at $z=0$ that is significantly closer to that from \citet{wheeler-19}, despite the differences in the cosmic ionising background.

As Figure~\ref{fig:MZR-UFD} shows, going from the original FIRE-2 version with the \citet{mannucci-06} DTD to the metallicity-independent DTD from \citet{maoz-17} to metallicity-dependent rate models increases the median iron abundance and in general reduces, but does not eliminate, the tension with observations. Considering only enriched stellar populations leads to much better agreement.
However, the differences across our metallicity-dependent models for Ia rates are too modest to provide a definitive test. We remind the reader that other simulation studies (EDGE, DC Justice League) also report the original discrepancy, and that a combination of modelling core-collapse supernova enrichment in Pop III stars and metallicity-dependent Ia rates might provide the best agreement in the stellar mass-metallicity relation at these galaxy masses.


\subsubsection{Stellar $\alpha$-element abundances}
\label{sec:3.2.5-MgFe-FeH}

\begin{figure*}
\centering
\begin{tabular}{c}
\includegraphics[width = 1.0 \linewidth]{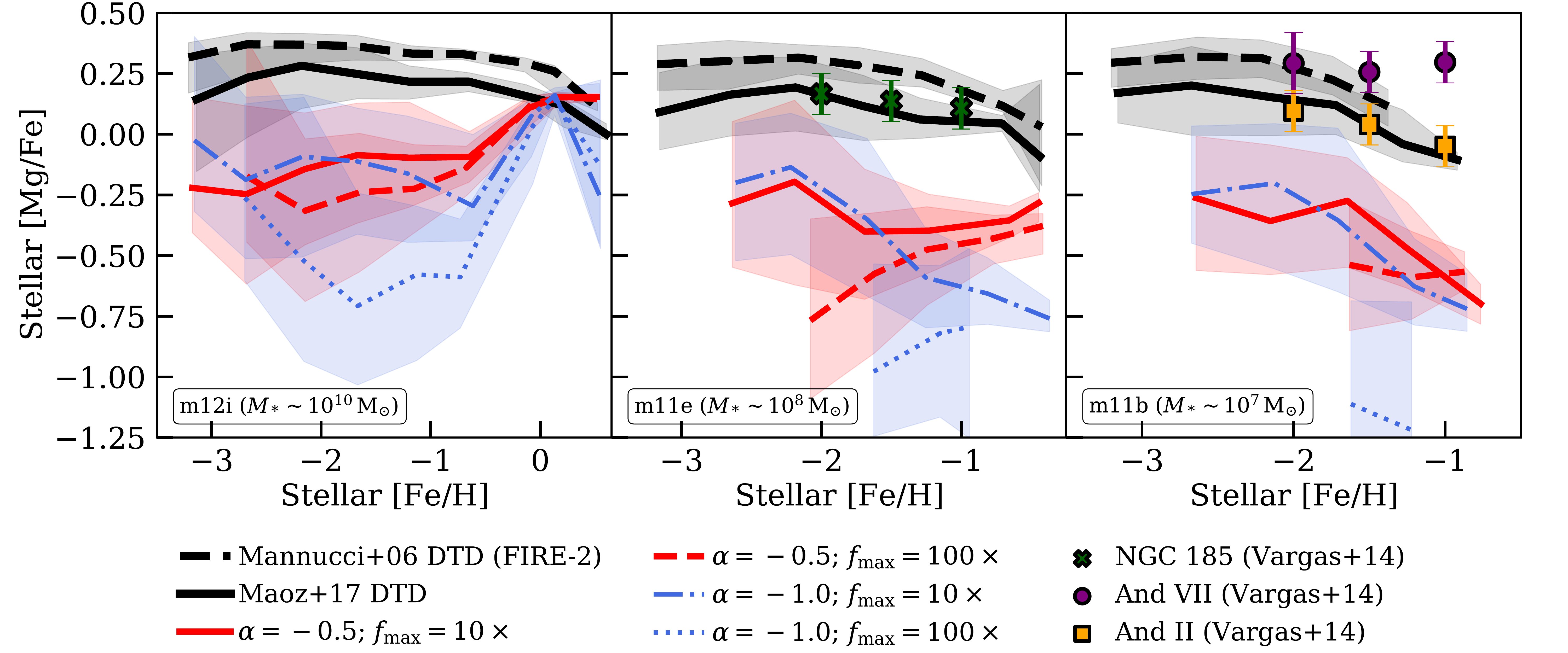}
\end{tabular}
\vspace{-2 mm}
\caption{
\textbf{$\alpha$-capture element abundances for (re)simulations with different models for Ia rates.}
We show the median (curves) as well as $68$ per cent scatter (shaded regions) in stellar [Mg/Fe] versus [Fe/H]. For the original FIRE-2 simulations, based on the Ia DTD from \citet{mannucci-06}, and those re-simulated with the (metallicity-independent) DTD from \citet{maoz-17}, we find that both the normalisation of [Mg/Fe] as well as the trend with [Fe/H] to be broadly consistent with observed abundances in Local Group galaxies \citep{vargas-14}. For re-simulations with metallicity-dependent Ia rates, we find a drop in the [Mg/Fe] normalisation by $\sim 0.5-0.75$ for models with $\alpha=-0.5$, and by $\sim 1.0-1.25$ for $\alpha=-1.0$. Some of the more extreme rate models lead to [Mg/Fe] increasing with [Fe/H] and an overall positive slope, which is inconsistent with current observations.
\textit{Overall, metallicity-dependent models for Ia rates lead to a shallower (and sometimes increasing) relation between [Mg/Fe] and [Fe/H], and more broadly, they reduce the overall [Mg/Fe] normalisation, enabling us to disfavour models with a strong dependence of Ia rate on $Z$ ($\alpha \sim -1.0$) or very large rate boost caps ($f_{\rm max} >> 10$).}
}
\label{fig:newruns-MgFe-FeH}
\end{figure*}


Finally, we investigate the impact of modifying rates of Ia supernovae on stellar alpha-to-iron ratios. Specifically, we consider stellar [Mg/Fe], because boosting the Ia rate without changing the core-collapse supernova rate should drive up iron abundances without changing $\alpha$-element abundances like magnesium, thus leading to lower [Mg/Fe].

We investigate stellar [Mg/Fe] versus [Fe/H] for 3 of our re-simulated galaxies in Figure~\ref{fig:newruns-MgFe-FeH}. For m11b and m11e, [Mg/Fe] versus [Fe/H] for the metallicity-independent DTDs from \citet{mannucci-06} and \citet{maoz-17} are broadly consistent with observations of select Local Group dwarf galaxies at similar masses from \citet{vargas-14}, as the right panel shows. For re-simulations with metallicity-dependent models for Ia rates, the median [Mg/Fe] drops as expected. These reductions in normalisation are $\sim 0.5-0.75$ for $\alpha=-0.5$ and $\sim 1-1.25$ for $\alpha=-1.0$. One caveat here is that uncertainties in our assumed core-collapse supernova rates and magnesium yields might also play a role in determining the overall [Mg/Fe] normalisation, but likely to a lesser extent that our strong metallicity-dependent Ia rate modifiers.

In the 2 lower-mass galaxies (m11e and m11b), we find for certain models that [Mg/Fe] values actually increase with [Fe/H], resulting in a positive slope, and a dearth of stars at low [Fe/H], given how rapidly [Fe/H] enriches at low [Fe/H] in the more extreme metallicity-dependent Ia models, especially during strong early burst of star formation.
The higher-mass simulation (m12i) also shows such positive slopes in some cases. This is possibly from early stellar populations being metal poor leading to increased Ia rates and a drop in [Mg/Fe], with later populations converging on solar (or higher) abundances and not having their Ia rates boosted significantly. Such an effect could result in the seen convergence of [Mg/Fe] values around [Fe/H] $\sim 0$. In either case, we are not aware of any observed significant positive slopes in [Mg/Fe] versus [Fe/H].

Figure~\ref{fig:newruns-MgFe-FeH} also shows, in the more extreme ($\alpha=-1.0$) rate variants of the two lower-mass galaxies m11e and m11b, a gap in [Fe/H] values, which `jump' to $\gtrsim -2.0$ and above. This is because both galaxies have relatively small numbers of enriched stars, and those enriched populations have abundances of [Fe/H] $\gtrsim -2.0$ from increased stellar feedback when their Ia rates were significantly boosted. In the extreme variants of both m11e and m11b, the greatly increased Ia feedback results in truncated, bursty star formation, and a majority of stars end up staying at the metallicity floor, while some enriched populations end up with [Fe/H] $\gtrsim -2.0$. This results in a narrow distribution of iron abundances for enriched stars at [Fe/H] $\approx -2$ to $-1$.

Overall, we find that the more extreme models for Ia rates (those with $\alpha << -0.5$ or $f_{\rm max} >> 10$) result in dramatic drops in stellar [Mg/Fe], or strongly positive trends in [Mg/Fe], versus [Fe/H], likely ruled out by observations. This allows us to disfavour those extreme models and instead favour ones with more modest dependence of Ia rate on overall metallicity.


\section{Summary and Conclusions}
\label{sec:4-discussion}

We explored a range of metallicity-dependent models for Ia supernova rates, motivated by the observed trends in specific Ia rate versus galaxy mass in surveys like ASAS-SN and DES, and by observations of the metallicity dependence of the close-binary fraction of Milky Way stars.
We also explored the impact of these models on various aspects of galaxy formation.

Some studies \citep[DES;][for example]{wiseman-21} argued that the trend in the specific Ia rate versus galaxy stellar mass is potentially consistent with a DTD similar to that of \citet{maoz-17}, without invoking any additional dependence. However, others \citep[such as ASAS-SN; ][]{brown-19} claim that such a DTD cannot be reconciled with observed trends. We find that differences between these works/surveys arise not primarily because of differences in Ia measurements or assumed DTDs, but rather, primarily from different assumed galaxy SMFs and SFHs at $z \sim 0$. If the low-mass SMF rises steeply \citep[as per][]{baldry-12}, late-time SFHs of low-mass galaxies would be higher and the inferred specific Ia rates for low-mass galaxies much lower -- in this case, the tension is significantly weaker but potentially still not reconciliable with metallicity-independent Ia rates. If the SMF is shallow at lower masses, with lower late-time SFHs \citep[as determined in][and predicted in the FIRE simulations]{bell-03}, then the tension is quite strong and additional metallicity dependence of the Ia DTD is definitely required.

\textit{Here we summarise our main results, including the various motivations for metallicity-dependent Ia rates, our range of explored models, and their impact on difference aspects of galaxy formation and evolution:}

\textbf{(i) Specific Ia rates depend mostly on recent SFHs and choice of SMFs, especially at the low-mass end:} We showed that for the DTD from \citet{maoz-17}, Ia rates at $z\sim0$ are mostly sensitive to recent star formation, $\lesssim 1$ Gyr. Additionally, assuming a steep low-mass SMF \citep[as in][]{baldry-12} and higher late-time SFRs results in a weaker inferred trend of Ia rate versus stellar mass, while a shallower low-mass SMF \citep[such as][]{bell-03} and lower late-time SFRs leads to a steeper dependence of Ia rate on galaxy mass.
The latter definitely cannot be reconciled with a vanilla DTD and we see weak tension even for the former, motivating additional dependence of Ia rates on a quantity like metallicity.

\textbf{(ii) Our metallicity-dependent model for Ia rates, motivated by observations:} Milky Way observations \citep{moe-19, wyse-20} of the close-binary fraction of Solar-type stars being anti-correlated with metallicity motivated our choice of metallicity-dependent modifiers to the Ia DTD from \citet{maoz-17}. We chose the form $\rm{Rate_{Ia}} \propto \rm{min[(\textit{Z/Z}_{\odot})^{\alpha}, f_{\rm max}]}$, with $-1 \leq \alpha \leq 0$ and $f_{\rm max} = 10\times$ or $100\times$. When computing rates in post-processing, we found a viable set of models that recover the full systematic spread in observed trends, including the steepest ones consistent with a shallower low-mass SMF.

\textbf{(iii) Re-simulations with self-consistent modifications to Ia rates improve agreement with observed rates except for when the more extreme cases lead to strongly increased feedback:} The re-simulated galaxies show minimal changes to stellar masses, sizes, circularity, and shapes, barring the most extreme models with $\alpha \lesssim -1.0$ or $f_{\rm max} \gtrsim 100$, which puff up, over-quench, and/or destroy their disk-like structure. The specific Ia rates in these re-simulations, while closer to the steeper observed trends, are at best $1-\sigma$ too shallow and do not resolve the tension completely because of the trade off between boosting supernova rates and forming fewer new stars overall.

\textbf{(iv) Better agreement with observed stellar mass-metallicity relation, with uncertainty for extremely low-mass galaxies:} While high-mass galaxies in the fiducial FIRE-2 suite show reasonable agreement with observed metallicities, lower-mass galaxies are systematically lower by $0.2 - 0.5$ dex \citep[as has also been shown previously by][]{ma-16, escala-18, hopkins-20}. The metallicity-dependent rate models increase the stellar iron abundance in simulations with lower stellar masses ($10^7\,\rm{M}_{\odot} < M_* <10^9\,\rm{M_{\odot}}$) and more metal-poor populations, improving somewhat the agreement with the observed mass-metallicity relation at low masses. For extremely low-mass ($M_* \sim 10^3 - 10^5\,\rm{M_{\odot}}$) galaxies, however, the self-limiting nature of increased Ia feedback (and possibly persistant resolution limits of our simulations) mean that that our modified models can account for only about half of the deficit in [Fe/H] compared with observations. Previously \citet{muley-21} have shown, using FIRE simulations, that age- and metallicity-dependent core-collapse supernova yields also do not solve the [Fe/H] discrepancy at extremely low masses. Both of these results point to potential new astrophysics or a combination of modifications to feedback and enrichment processes being the solution, motivating future work on this subject that explores different factors in combination.

\textbf{(v) Impact of modifying Ia rates on alpha element abundances:} The primary effect of boosting Ia rates is to lower [Mg/Fe], with a higher-order effect altering the slope of the [Mg/Fe] versus [Fe/H] trend, especially in low-mass galaxies. This enables us to strongly disfavour our more extreme models ($\alpha << -0.5$ and/or $f_{\rm max}>>10$) in favour of those with a more moderate dependence of Ia rates on metallicity.

Although Type Ia supernovae are only one piece of the puzzle for understanding stellar populations and modelling feedback in galaxies (along with core-collapse supernovae, stellar winds, photoionisation, photoelectric heating, etc.), understanding their numbers, rates, and energetics is nonetheless crucial to a wide variety of topics in astrophysics and cosmology. \textit{To our knowledge, this work is the first to explore metallicity-dependent Ia rates and their impact using cosmological simulations of galaxy formation. We hope to lay the groundwork for and motivate future studies in this area.}


\section*{Acknowledgements}

We thank Peter Behroozi, Ivanna Escala, Shea Garrison-Kimmel, Robyn Sanderson, Dan Weisz, and Philip Wiseman for valuable discussions that improved this paper overall, as well as sharing data in some cases.

This analysis relied on \textsc{NumPy} \citep{numpy-20}, \textsc{SciPy} \citep{scipy-01, scipy-20}, \textsc{AstroPy}\footnote{\url{https://www.astropy.org/}}, a community-developed core Python package for Astronomy \citep{astropy-13, astropy-18}, \textsc{Matplotlib}, a Python library for publication-quality graphics \citep{matplotlib-07}, the \textsc{IPython} package \citep{ipython-07}, and the publicly available package \textsc{GizmoAnalysis} \citep[][available at \url{https://bitbucket.org/awetzel/gizmo\_analysis}]{wetzel-20}; as well as NASA's Astrophysics Data System (ADS)\footnote{\url{https://ui.adsabs.harvard.edu/}} and the \textsc{arXiv}\footnote{\url{https://www.arxiv.org/}} preprint service.

PJG and AW received support from: the NSF via CAREER award AST-2045928 and grant AST-2107772; NASA ATP grants 80NSSC18K1097 and 80NSSC20K0513; HST grants GO-14734, AR-15057, AR-15809, and GO-15902 from STScI; a Scialog Award from the Heising-Simons Foundation; and a Hellman Fellowship. AW and BJS acknowledge the Scialog Fellows program, sponsored by the Research Corporation for Science Advancement, which motivated some of this work. Support for PFH was provided by NSF Research Grants 1911233 \&\ 20009234, NSF CAREER grant 1455342, NASA grants 80NSSC18K0562, HST-AR-15800.001-A. BJS received support from NSF grants AST-1920392,  AST-1911074,  AST-1908952, and AST-2050710 and NASA grants HST-GO-16451, HST-GO-16498, and 80NSSC21K1788. CW acknowledges support from NSF LEAPS-MPS grant AST-2137988. CAFG received support from NSF grants AST-1715216, AST-2108230,  and CAREER award AST-1652522; from NASA grant 17-ATP17-0067; from STScI through grant HST-AR-16124.001-A; and from the Research Corporation for Science Advancement through a Cottrell Scholar Award. 

We ran simulations and performed numerical calculations using: the UC Davis computer cluster Peloton, the Caltech computer cluster Wheeler, the Northwestern computer cluster Quest; XSEDE, supported by NSF grant ACI-1548562; Blue Waters, supported by the NSF; Frontera allocations FTA/Hopkins-AST21010 and AST20016, supported by the NSF and TACC; XSEDE allocations TG-AST140023 and TG-AST140064, and NASA HEC allocations SMD-16-7561, SMD-17-1204, and SMD-16-7592; Pleiades, via the NASA HEC program through the NAS Division at Ames Research Center.

\section*{Data Availability}

The python code and data tables used to create each figure are available at \url{https://github.com/pratikgandhi95/Ia-rates-metallicity-dependence}. The FIRE-2 simulations are publicly available \citep{wetzel-22} at \url{http://flathub.flatironinstitute.org/fire}. Additional FIRE simulation data is available at \url{https://fire.northwestern.edu/data/}. A public version of the GIZMO code is available at \url{http://www.tapir.caltech.edu/~phopkins/Site/GIZMO.html}.



\bibliographystyle{mnras}
\bibliography{SNrates} 




\appendix

\section{Sensitivity of Ia rates to stellar ages}
\label{app:A}

\begin{figure}
\centering
\begin{tabular}{c}
\includegraphics[width = 0.98 \linewidth]{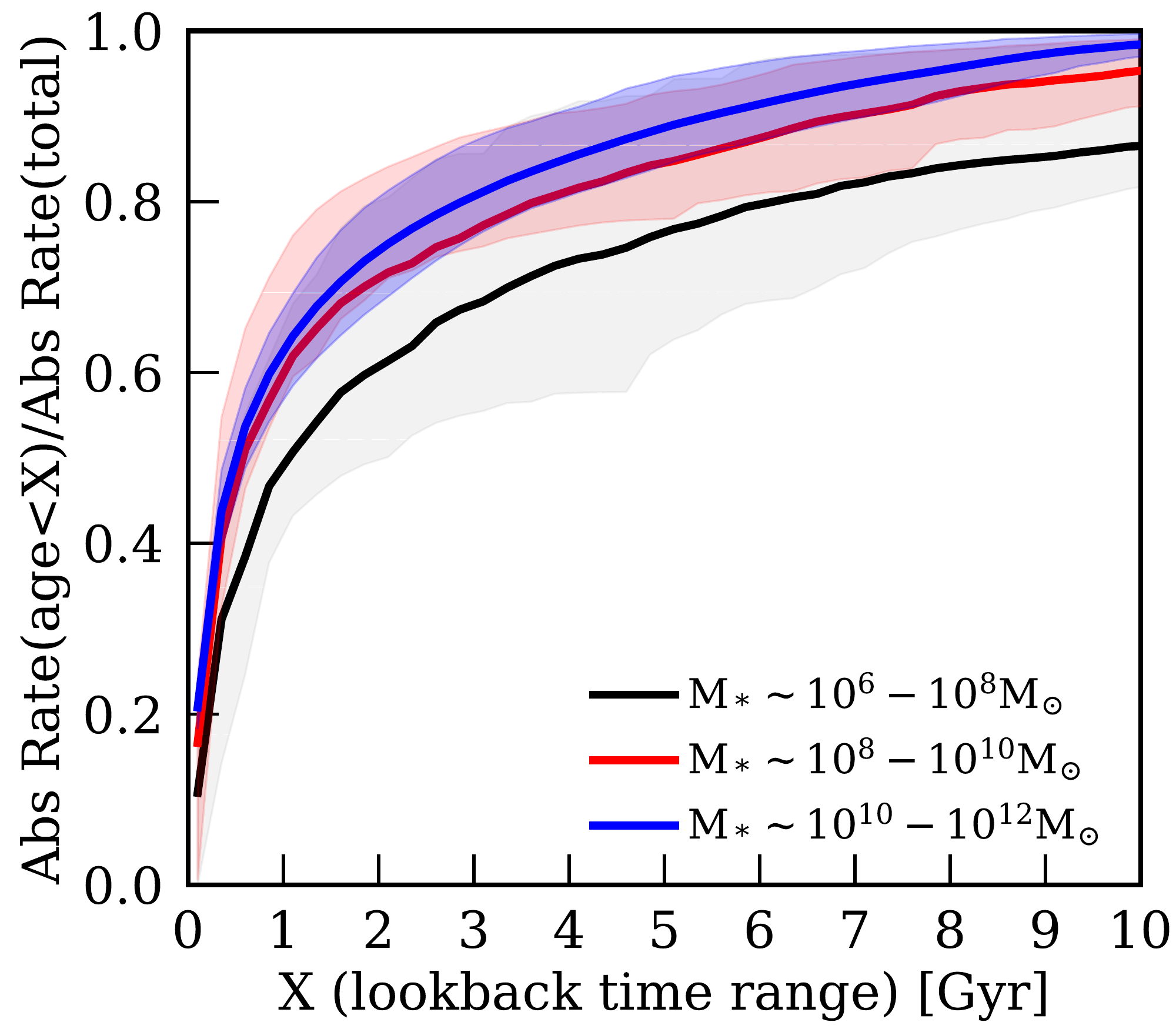}
\end{tabular}
\vspace{-2 mm}
\caption{
\textbf{Dependence of Ia rates at $z=0$ on stellar age in the FIRE-2 simulations}.
For the (metallicity-independent) delay time distribution from \citet{maoz-17} (re-computed in post-processing), we show the cumulative contribution fraction to the total Ia rate at $z = 0$ from stellar populations younger than a given age. The 3 lines and shaded regions show simulations grouped into 3 stellar mass bins. At each age, we show the median and $68$ per cent spread of this fraction across the simulations. Considering only stars younger than $\sim 1.1$ Gyr, $0.75$ Gyr, and $0.63$ Gyr accounts for the majority of the $z=0$ Ia rate for galaxies in stellar mass ranges of $10^6$ - $10^8 \, \rm{M}_{\odot}$, $10^8$ - $10^{10} \, \rm{M}_{\odot}$, and $10^{10}$ - $10^{12} \, \rm{M}_{\odot}$ respectively.
\textit{For star-forming galaxies, Ia rates are sensitive primarily to only recent ($\lesssim 1$ Gyr) star formation, which motivates our benchmarks of recent SFHs in Figures~\ref{fig:SFMS-z0} and \ref{fig:tau-90}.}
}
\label{fig:rate-lbt}
\end{figure}

Here, we further explore the sensitivity of Ia rates on the galaxy SFH and ages of the stellar populations, using the \citet{maoz-17} DTD applied to our FIRE-2 galaxies.
Figure~\ref{fig:rate-lbt} shows, for bins in present-day stellar mass, the cumulative contribution fraction to the total present-day Ia rate in our simulations from stellar populations younger than a given age (measured in lookback time from $z=0$). By considering only those stellar populations formed within different ranges in lookback time from present day, we compute the ratio of absolute Ia rates from the selected stars relative to the total Ia rate from all stars. Including only stellar populations formed within the last $\sim 1$ Gyr accounts for the majority of the present-day Ia rates across all galaxy masses that we examine.

\textit{Thus, for all star-forming galaxies and simulations in our analysis, the Ia rates are sensitive primarily to recent ($\lesssim 1$ Gyr) star formation.} This motivates why, in Section~\ref{sec:2.2-sfh-benchmarking} and Appendix~\ref{app:B}, we benchmark our simulated SFHs to observational and semi-empirical results using metrics for only recent SFHs: the $90$ per cent stellar mass assembly timescales in Figure~\ref{fig:tau-90} and the star-formation rates at $z\sim0$ in Figure~\ref{fig:SFMS-z0}.

\section{Additional benchmarks of simulated star formation histories}
\label{app:B}

\begin{figure}
\centering
\begin{tabular}{c}
\includegraphics[width=0.98\linewidth]{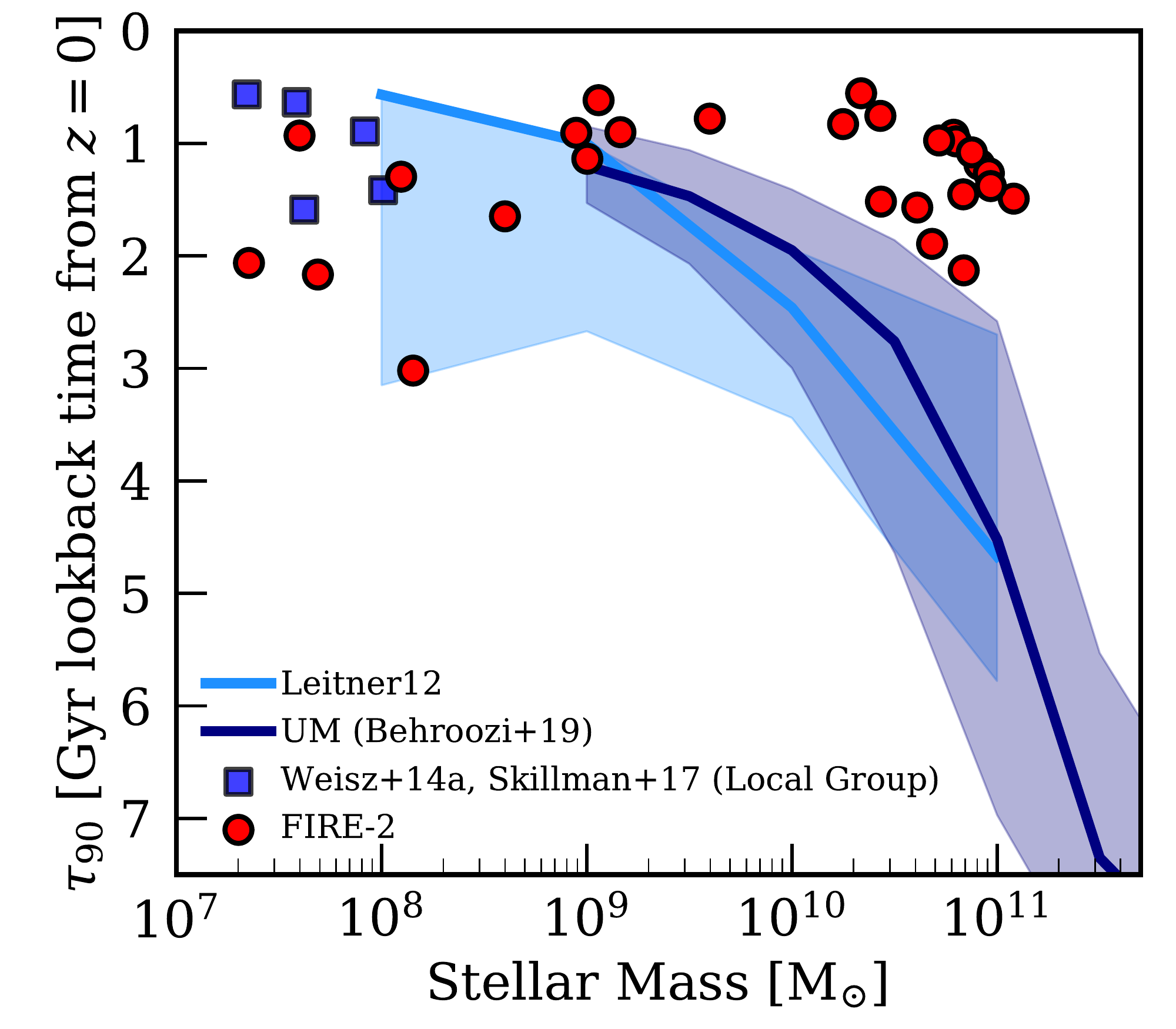}
\end{tabular}
\caption{
\textbf{Lookback time of the assembly of $90$ per cent of stellar mass at $z=0$.} $\tau_{90}$ for \textsc{FIRE-2} galaxies (red points), compared with observations of star-forming galaxies in the Local Group from \citet{weisz-14} and \citet{skillman-17} (blue points), observations of star-forming galaxies from \citet{leitner-12} (blue curve with $68$ per cent scatter in shaded region), and from the \textsc{UniverseMachine} (UM) semi-empirical model \citep[][navy curve with $68$ per cent scatter in shaded region]{behroozi-19}. \textit{FIRE-2 galaxies with stellar masses between $10^7 \rm{M}_{\odot}$ and $10^9 \rm{M}_{\odot}$ agree with observed and semi-empirical $\tau_{90}$ to within $1 \sigma$. At higher masses, \textsc{FIRE}-2 galaxies assemble later than semi-empirical constraints, with $> 1 \sigma$ disagreement.} 
}
\label{fig:tau-90}
\end{figure}

We present another benchmark of the SFHs of FIRE-2 simulations, especially for our lower-mass galaxies, beyond the sSFRs in Figure~\ref{fig:SFMS-z0}. We consider $\tau_{90}$, the lookback time prior present day when a galaxy assembled $90$ per cent of its current stellar mass. For reference, for a simple flat SFH (constant SFR), $\sim 70$ per cent of the present-day Ia rate comes from stars that form after $\tau_{90}$. Figure~\ref{fig:tau-90} shows $\tau_{90}$ for our fiducial suite of FIRE-2 galaxies. For comparison, we show $\tau_{90}$ values from observations compiled by \citet{leitner-12}, the semi-empirical model \textsc{UniverseMachine} \citep[UM;][]{behroozi-19}, as well as those for Local Group galaxies \citep[][]{weisz-14, skillman-17}. For galaxies with $M_* \lesssim 10^9 \, \rm{M}_{\odot}$, we find good agreement to within $1\sigma$ between our simulations and observations. At higher masses, we find that FIRE-2 galaxies systematically form later than semi-empirical models would suggest, with $> 1\sigma$ disagreement. A previous comparison of stellar mass assembly of $M_* \geq 10^{10}\,\rm{M}_{\odot}$ mass galaxies in FIRE in \citet{santistevan-20} appears to show better agreement with the \citet{leitner-12} and UM trends, but we note that their comparison in their Figure~2 is in redshift space, which likely causes the discrepancy to appear smaller than it does in linear time units here.

For higher mass galaxies, because Ia rates are primarily sensitive to a galaxy's recent ($\lesssim 1$ Gyr) star-formation, the sSFR comparison in Figure~\ref{fig:SFMS-z0} is more relevant than $\tau_{90}$. However, for $M_* \lesssim 10^9\,\rm{M}_{\odot}$, star formation on a slightly longer timescale matters, so this agreement of $\tau_{90}$ values with observations lends confidence to the Ia rates in our low-mass FIRE-2 galaxies - with the still important note of caution that our stellar mass dependence of both sSFR and specific Ia rates in FIRE-2 may be shallower than observed (as per Figures \ref{fig:SFMS-z0} and \ref{fig:nomet-rates}).


\bsp	
\label{lastpage}
\end{document}